\documentclass{aa}  

\usepackage{graphicx}
\usepackage{natbib}
\usepackage{txfonts}
\newcommand{\kep}{{\it Kepler}}
\newcommand{\prot}{$P_\text{rot}$}
\newcommand{\sph}{$S_\text{\!ph}$}
\newcommand{\teff}{$T_\text{eff}$}
\newcommand{\logg}{$\log\, g$}

\newcommand{\tobs}{$t_\text{obs}$}
\newcommand{\F}{$F_{10.7}$}
\usepackage{colortbl}
\newcommand{\avsph}{$\langle S_\text{\!ph}\rangle$}
\newcommand{\stdsph}{$\sigma(S_\text{\!ph})$}
\newcommand{\metal}{[Fe/H]}

\let\OLDthebibliography\thebibliography
\renewcommand\thebibliography[1]{
  \OLDthebibliography{#1}
  \setlength{\parskip}{0pt}
  \setlength{\itemsep}{0pt plus 0.3ex}
}

\definecolor{purple}{RGB}{102,0,204}
\definecolor{blue2}{RGB}{29,95,161}
\definecolor{orange}{RGB}{225,130,0}
\definecolor{green2}{RGB}{29,161,95}
\definecolor{greenorcid}{RGB}{166,206,57}

\usepackage{hyperref,multirow}
\hypersetup{colorlinks,citecolor={blue2},linkcolor={blue2},urlcolor={blue2}}  

\begin{document}

   \title{Temporal variation of the photometric magnetic activity for the Sun and \textit{Kepler} solar-like stars} 

   \author{A. R. G. Santos\inst{1,2} %0000-0001-7195-6542
          \and S. Mathur \inst{3,4} %0000-0002-0129-0316
          \and R. A. Garc\'{i}a \inst{5} %0000-0002-8854-3
          \and A.-M. Broomhall\inst{2} %0000-0002-5209-9378 
          \and R. Egeland \inst{6} %0000-0002-4996-0753
          \and A. Jim\'enez \inst{3,4}
          \and D. Godoy-Rivera \inst{3,4}
          %0000-0003-4556-1277
          \and S. N. Breton \inst{7,8} %0000-0003-0377-0740
          \and Z. R. Claytor \inst{9,10} %0000-0002-9879-3904
          \and T. S. Metcalfe \inst{11} %0000-0003-4034-0416
          \and M. S. Cunha \inst{1}
          %0000-0001-8237-7343
          \and L. Amard \inst{5} %0000-0003-2508-6093
          }

   \institute{Instituto de Astrof\'isica e Ci\^encias do Espa\c{c}o, Universidade do Porto, CAUP, Rua das Estrelas, PT4150-762 Porto, Portugal \\
            \email{Angela.Santos@astro.up.pt}
         \and
             Department of Physics, University of Warwick, Coventry, CV4 7AL, UK
         \and
             Instituto de Astrof\'\i sica de Canarias (IAC), E-38205 La Laguna, Tenerife, Spain
         \and
             Universidad de La Laguna (ULL), Departamento de Astrof\'isica, E-38206 La Laguna, Tenerife, Spain
         \and
             Universit\'e Paris-Saclay, Universit\'e Paris Cit\'e, CEA, CNRS, AIM, 91191, Gif-sur-Yvette, France
         \and
             High Altitude Observatory, National Center for Atmospheric Research, P.O. Box 3000, Boulder, CO 80307-3000, USA
         \and
             Universit\'e Paris Cit\'e, Universit\'e Paris-Saclay, CEA, CNRS, AIM, 91191, Gif-sur-Yvette, France
         \and
             INAF – Osservatorio Astrofisico di Catania, Via S. Sofia, 78, 95123 Catania, Italy
         \and
             Department of Astronomy, University of Florida, 211 Bryant Space Science Center, Gainesville, FL 32611, USA
         \and
             Institute for Astronomy, University of Hawai'i at M\={a}noa, 2680 Woodlawn Drive, Honolulu, Hawai'i 96822, USA
         \and
             White Dwarf Research Corporation, 9020 Brumm Trail, Golden, CO 80403, USA
         %\and
         %   Space Science Institute, 4765 Walnut St., Suite B, Boulder, CO 80301, USA
          }

   \date{\today}
   
   \titlerunning{\sph\ variability for the Sun and solar-like stars}
   
   \authorrunning{A. R. G. Santos}

% \abstract{}{}{}{}{} 
% 5 {} token are mandatory
 
  \abstract
   {The photometric time series of solar-like stars can exhibit rotational modulation, that is, brightness variations due to active regions co-rotating with the stellar surface. These signatures allow us to constrain properties of stellar rotation and magnetic activity.}
   {In this work we investigate the behavior, particularly the variability in terms of strength, of the photometric magnetic activity of \kep\ solar-like stars and compare it with that of the Sun. }%Our goal is also to understand the impact of the limited observational time span.}
   {We adopted the photometric magnetic activity proxy, \sph, which was computed with a cadence of five times the rotation period (\prot). The average \sph\ was taken as the mean activity level, and the standard deviation was taken as a measure of the temporal variation of the magnetic activity over the \kep\ observations. We also analyzed Sun-as-a-star photometric data from VIRGO (Variability of Solar Irradiance and Gravity Oscillations). Sun-like stars were selected from a very narrow parameter space around the solar properties, according to the recent \textit{Gaia-Kepler} stellar properties catalog and the latest \kep\ rotation catalog. We also looked into KIC~8006161 (HD~173701), a very active metal-rich G dwarf, and we compared its magnetic activity to that of stars with similar stellar fundamental parameters.}
   {We find that the amplitude of \sph\ variability is strongly correlated with its mean value, independent of spectral type. An equivalent relationship has previously been found for ground-based observations of chromospheric activity emission and magnetic field strength, but in this work we show that photometric \kep\ data also present the same behavior. While, depending on the phase of the cycle, the Sun is among the less active stars, we find that the $\text{S}_{\text{ph}\odot}$ properties are consistent with those observed in \kep\ Sun-like stars. KIC 8006161 is, however, among the most active of its peers, which tend to be metal-rich. This results from an underlying relationship between \prot\ and metallicity and supports the following interpretation of the magnetic activity of KIC 8006161:\  its strong activity is a consequence of its high metallicity, which affects the depth of the convection zone and, consequently, the efficiency of the dynamo. }
   {}


   \maketitle
%
%-------------------------------------------------------------------

\section{Introduction}\label{sec:intro}

Low-mass stars with convective envelopes, particularly stars of spectral type FGKM (hereafter solar-like stars), can be magnetically active \citep[e.g.,][]{Brun2017}. As with the Sun, the emergence of dark magnetic spots at the surface is one of the manifestations of their magnetic activity. As stars rotate, the dark starspots go in and out of view while modulating the stellar brightness. Bright faculae in active regions can also produce rotational modulation; however, their intensity contrast is smaller. Hence, the rotational modulation contains information on stellar surface rotation and magnetic activity. Both of these properties are strongly dependent on stellar age \citep[e.g.,][]{Wilson1963,Wilson1964,Skumanich1972,Kawaler1988}. In particular, during their main sequence, as stars evolve and lose angular momentum due to magnetic braking, they spin down following the so-called Skumanich law \citep{Skumanich1972}. The possibility of constraining stellar ages through surface rotation measurements led to the formulation of gyrochronology \citep{Barnes2003}, which is also mass dependent. After solar-like stars converge to a narrow rotation sequence, lower-mass stars spin down faster than higher-mass stars \citep[e.g.,][]{Barnes2003,Barnes2007,vanSaders2013,Matt2015}. However, stars above the Kraft break \citep[$T_\text{eff}>6250$ K;][]{Kraft1967}, with very shallow convective envelopes, are unable to spin down significantly.

Space-based photometric planet-hunting missions provide a unique opportunity to characterize stars. Namely, the \kep\ main mission \citep{Borucki2010} obtained unmatched long-term and continuous observations for hundreds of thousands of stars \citep[e.g.,][]{Mathur2017}. For those exhibiting rotational modulation (i.e., several tens of thousands of solar-like stars), \kep\ data allowed average surface rotation periods (\prot) to be measured \citep[e.g.,][]{Nielsen2013,Reinhold2013a,McQuillan2013a,McQuillan2014,Garcia2014,Ceillier2016,Ceillier2017,Santos2019a,Santos2021ApJS,Breton2021}. For main-sequence \kep\ solar-like stars, surface rotation periods generally decrease with increasing effective temperature: M stars have a median period of $\sim40$ days, while for F stars the median period is $\sim 10$ days. This general dependence is consistent with the expected rotation evolution \citep[e.g.,][]{Barnes2003,vanSaders2013,Matt2015} and also with the narrow rotation sequences in cluster data \citep[e.g.,][]{Barnes2003,Agueros2011,Rebull2016,Fritzewski2021,Boyle2023}. Interestingly, for \kep\ GKM main-sequence (i.e., dwarfs) stars, which span a relatively wide range of ages, the rotation-period distribution is bimodal and stars tend to group into two populations or groups. The two populations are separated by an intermediate-\prot\ gap.

When first discovered, the bimodal rotation-period distribution was thought to be a feature of the \kep\ field, and its origin was attributed to two distinct epochs of star formation \citep[][see also \citealt{Davenport2018}]{McQuillan2013a,McQuillan2014}. However, it is now known that the rotation-period distribution is also bimodal across the different fields of view of the extended \kep\ mission, K2 \citep{Howell2014}, as reported by \citet{Reinhold2020} and \citet{Gordon2021}. Two other hypotheses have been suggested to explain this phenomenon. \citet{Montet2017} and \citet{Reinhold2019} find evidence for the stars in the fast-rotating population being spot-dominated and those in the slow-rotating population being faculae-dominated. In this context, stars would transition from spot- to faculae-dominated and the ``gap'' in the rotation distribution would be an artifact resulting from a lack of rotation detections due to the cancelation between the dark spots and the bright faculae \citep{Reinhold2019}. Another interpretation of the bimodal rotation distribution has also been proposed: a broken spin-down law \citep{McQuillan2014,Angus2020,Spada2020,Gordon2021,Lu2022}. While decoupled from the core, the surface of stars on the fast-rotating population spins down due to magnetic braking, following the Skumanich law. As the fast-rotating core and the envelope start to couple, the surface spin-down would cease. This transition would be relatively fast, leading to the intermediate-\prot\ gap. Having completed the coupling (reaching the slow-rotating population), the stellar surface would start to spin down again following the Skumanich law. Recently, evidence supporting this interpretation of the intermediate-\prot\ gap was found in ground-based photometric data by \citet{Lu2022}. The authors find that the gap is absent in the regime of fully convective stars, while it is still found for the partially convective stars.

In addition to surface rotation, spot modulation also allows stellar magnetic activity to be constrained. Despite the degeneracy between different parameters, the amplitude of the rotational signal in the light curve is related to the spot coverage of the stellar surface, which in turn is related to the magnetic activity level of the star. Based on this and developing upon the starspot proxy used in \citet{Garcia2010}, \citet{Mathur2014} defined the photometric magnetic activity index, \sph. It is a measurement of the amplitude of the rotational modulation and has been shown to be an adequate proxy of solar and stellar magnetic activity \citep{Salabert2016a,Salabert2017}. For the Sun, \citet{Salabert2017} demonstrated that the \sph\ is well correlated with other, more conventional, proxies of magnetic activity, such as the sunspot number, sunspot areas (SAs), Ca H-K emission, and radio flux. Based on seismic solar analogs observed by \kep, \sph\ was also found to be complementary to the Ca H-K emission for stars other than the Sun \citep[][see also {\color{blue2}Egeland et al. in prep.}]{Salabert2016a,Karoff2018}. For \kep\ main-sequence solar-like stars, the dependence of \sph\ on effective temperature is complex \citep{Santos2019a,Santos2021ApJS}. M stars tend to have large \sph; the \sph\ slightly decreases toward late-type K stars. From $\sim 4000$ K to $\sim 6000$ K, the range of \sph\ widens: the upper edge of the \sph\ distribution moves toward larger \sph\ values, while the lower edge moves toward smaller values and changes more drastically. F stars typically have small \sph, which may be attributed to their shallow convective envelopes and to their fast evolution, which also may contribute to short activity lifetimes \citep[e.g.,][]{Reiners2012}.

Generally, fast-rotating stars are expected to have a stronger magnetic activity in comparison with slow-rotating stars as a result of the efficiency of the dynamo. Indeed, such a relationship has been observed in chromospheric and coronal magnetic activity proxies \citep[APs; e.g.,][]{Vaughan1981,Baliunas1983,Noyes1984b,Soderblom1993,Wright2011}. Similarly, for the \kep\ solar-like stars, the average photometric AP, \sph, tends to increase with decreasing rotation period \citep{Santos2019a,Santos2021ApJS}. However, the activity-rotation relationship is not linear. In fact, because of the \prot\ bimodality and intermediate-\prot\ gap in the \kep\ field mentioned above, the \sph\ versus \prot\ diagram exhibits two different regimes: for instance, for K stars there are two almost parallel populations. While the \prot\ distribution of F stars is not bimodal, there are also two groups of stars with different behaviors: those below the Kraft break and those above \citep[see Appendix~B in][]{Santos2021ApJS}. For late F stars below the Kraft break ($T_\text{eff}<6250$ K), \prot\ and \sph\ are well correlated, similar to what is found for cooler solar-like stars.
For stars above the Kraft break ($T_\text{eff}>6250$ K), the correlation between \prot\ and \sph\ is very weak, and a group of low-activity fast rotators emerges. As mentioned above, their thin convective envelopes may not be able to harbor a strong surface magnetic activity; therefore, these stars do not brake significantly along the main sequence either. Furthermore, the \kep\ F-star sample includes stars of different absolute and relative ages, and evolved F stars have been found to be among the most inactive stars in other stellar samples \citep[e.g.,][]{Wright2004,Schroder2013}.

In addition to the intermediate-\prot\ gap and the Kraft break, other transitions in the activity-rotation relationship have been reported for solar-like stars. Particularly, at activity levels consistent with the slow-rotating population, that is, stars with longer \prot\ than the intermediate-\prot\  gap, there is a lack of stars with intermediate Ca H-K emission, which is known as the Vaughan-Preston (VP) gap \citep[e.g.,][]{VaughanPreston1980,Vaughan1980,Henry1996,GomesDaSilva2021}. However, so far, such a gap has not been found in the photometric magnetic activity of \kep\ stars.

In this work we investigate the temporal variability of the photometric magnetic AP, \sph, over the \kep\ observations and compare the \sph\ behavior with the photometric magnetic activity of the Sun. The paper is organized as follows. We summarize the data and the target sample of this work in Sect.~\ref{sec:datasample}. Section~\ref{sec:sph} briefly describes the adopted magnetic AP, \sph. In Sect.~\ref{sec:results} we show our results for the Sun and solar-like stars. Moreover, by selecting stars very similar to the Sun and to KIC~8006161, a metal-rich solar analog with a known magnetic activity cycle, we investigate how the behavior of these stars matches that of the others. Section~\ref{sec:conclusion} presents a further discussion and our conclusions. 

\section{Data preparation and sample selection}\label{sec:datasample}

\subsection{Data preparation}\label{sec:data}

In this work, we analyze the KEPSEISMIC\footnote{MAST: \url{https://doi.org/10.17909/t9-mrpw-gc07}} data products obtained from \kep\ long-cadence data \citep[$\Delta t =29.42$ min;][]{Borucki2010} with the \kep\ Asteroseismic Data Analysis and Calibration Software  \citep[KADACS;][]{Garcia2011}. KEPSEISMIC data were optimized for seismic studies and are also proper for the analysis of spot modulation in the light curves \citep[][]{Santos2019a,Santos2021ApJS}. KEPSEISMIC data are obtained with customized apertures and corrected for outliers, jumps, drifts, and discontinuities at the \kep\ Quarter edges. In addition, gaps smaller than 20 days were filled using in-painting techniques \citep[see][]{Garcia2014a,Pires2015}. Finally, the light curves were high-pass filtered at 20, 55, and 80 days (i.e., three KEPSEISMIC light curves were analyzed per star). While the filters with short cutoff period are more efficient at minimizing the \kep\ instrumental effects, they can also filter the long-period stellar signal. Therefore, \citet{Santos2019a,Santos2021ApJS} conducted a parallel analysis of the different filtered light curves in order to find the best compromise between filtering out the instrumental modulations and preserving the rotational signal. Filtering the instrumental modulation is also useful to reduce the bias on the photometric magnetic AP, \sph. The choice for the most adequate filter for each target was already made in \citet{Santos2019a,Santos2021ApJS}. We note that periods longer than the cut-off period of a given filter employed by KADACS can still be retrieved, as the filter transfer function slowly reaches zero at twice the cut-off period.

Average rotation period and average \sph\ are adopted from \citet{Santos2019a,Santos2021ApJS}.
The rotation pipeline used by the authors employs three rotation diagnostics: wavelet analysis, autocorrelation function (ACF), and composite spectrum \citep[CS; e.g.,][]{Mathur2010b,Garcia2014,Ceillier2016,Ceillier2017}. The latter diagnostic combines the former two, highlighting the common peaks and attenuating the remainder, which is particularly relevant to avoid false positives due to, for example, instrumental effects or high-amplitude second and third harmonics. For each target there are nine period estimates (three per light curve). The final rotation periods were selected in three steps: automatic selection, which requires agreement between different estimates and a minimum height of the corresponding rotation peaks; machine learning through the implementation of ROOSTER \citep[Random fOrest Over STEllar Rotation;][]{Breton2021}, which, after an adequate training, was fed with stellar parameters, the output parameters from the rotation pipeline, and the associated \sph\ values; and complementary visual inspection. In all steps, the priority was given to the period-estimate from the wavelet analysis, which allows for a conservative approach, providing associated uncertainties of about 10\% in average. Further details are available in \citet{Santos2019a,Santos2021ApJS} and \citet{Breton2021}.

For the stellar effective temperature (\teff), surface gravity (\logg), and metallicity (\metal), we prioritized spectroscopic constraints when available; otherwise, we resorted to photometric values. The order of prioritization is as follows: \kep\ Community Follow-up Observation Program (CFOP) in high resolution  \citep[][]{Furlan2018}; Apache Point Observatory for Galactic Evolution Experiment \citep[APOGEE;][]{Ahumada2020}; Large Sky Area Multi-Object Fiber Spectroscopic Telescope \citep[LAMOST;][]{Zhao2012,Zong2020}; the \textit{Gaia}-\kep\ Stellar Properties Catalog \citep[GKSPC;][]{Berger2020}; and \kep\ Stellar Properties Catalog for Data Release 25  \citep[KSPC DR25;][]{Mathur2017}. We also adopted the model luminosities ($L$) and equivalent evolutionary phases (EEPs) obtained in {\color{blue2}Mathur et al. (in prep.)} using the implementation of the Yale Rotation Evolution Code by \citet{vanSaders2013} and \citet{vanSaders2016} and the interpolation tool by \citet[][\texttt{kiauhoku}]{Claytor2020,Claytor2020b}. Finally, in Appendix~\ref{sec:cleansample} we also adopt the renormalized unit-weight error \citep[RUWE;][]{GaiaDR1,GaiaEDR3,Lindegren2021}.

To place the Sun in context of solar-like stars, we used more than 24 years of data from the Variability of Solar Irradiance and Gravity Oscillations (VIRGO) sun photometer (SPM), which is on board the Solar and Heliospheric Observatory \citep[SoHO;][]{Domingo1995,Frohlich1995,Frohlich1997,Jimenez2002}, covering two full solar cycles: starting on 23 January 1996 and ending on 15 May 2021. For the final solar light curve, we only combined the {\sc green} and {\sc red} channels (hereafter VIRGO g+r), which best represent the \kep\ bandwidth \citep{Basri2010}. VIRGO/SPM collects observations with a cadence of 60 seconds, and thus we re-binned the data to 30 minutes to be consistent with \kep\ long-cadence data and reduce computing time (note that a long cadence is sufficient to study the rotational modulation of \kep\ solar-like stars, whose characteristic timescales are relatively long).

In Appendix~\ref{sec:appsun} we also use the daily records for the SAs\footnote{\url{solarcyclescience.com}} and for the solar flux at 10.7 cm\footnote{\url{www.ngdc.noaa.gov}} (\F). Sunspot areas are expressed in millionths of a solar hemisphere ($\mu\text{Hem}$), while \F\ is expressed in solar flux units (sfu; i.e., $10^{22}\, \text{W}\,\text{m}^{-2}\,\text{Hz}^{-1}$). The SA and \F\ records are available since 1 May 1874 and 17 February 1947, respectively. Sunspot areas, which correspond to the total sunspot coverage of the solar disk, are directly linked to the strong component of the magnetic field in the photosphere. \F\ is sensitive to both strong and weak components of the solar magnetic field in the upper chromosphere and lower corona \citep[e.g.,][]{Covington1969,Tapping1987,Tapping1990}.

\subsection{Sample selection}\label{sec:sample}

We started with the 55,232 stars of spectral type mid-F to M from \citet{Santos2019a,Santos2021ApJS}. Potential contaminants, such as eclipsing binaries, misclassified red giants, and classical pulsators, were discarded by the authors for the rotational analysis.

To further avoid biases, in this work, we also neglected the targets that required particular care to retrieve the correct rotation period due to the presence of high-amplitude instrumental modulation. Those were identified and corrected during the visual inspection in \citet{Santos2019a,Santos2021ApJS}. We also removed the targets for which we had to consider an individual photon-shot-noise correction \citep[see Sect.~\ref{sec:sph} and][]{Santos2019a,Santos2021ApJS} to ensure that the \sph\ values are computed in a homogeneous way. These two criteria removed 2,758 targets.

We also removed from the sample targets that have been identified as potential binaries: \textit{Gaia} DR2 and DR3 binaries \citep[][2,305 targets]{Berger2018,Binaries_GaiaDR3}, tidally synchronized binaries \citep[][130 targets]{Simonian2019}, and close-in binary candidates \citep[][2,486]{Santos2019a,Santos2021ApJS}. The respective flags were provided in \citet{Santos2019a,Santos2021ApJS}. 
The quasi-periodic modulation seen in close-in binary candidates identified by \citet{Santos2019a,Santos2021ApJS} can still be related to rotation. However, there are some striking differences between their signal and the typical rotation signature of solar-like stars. Close-in binary candidates have very large average \sph\ values, fast and stable beating patterns, and a long chain of high-amplitude harmonics. As shown in \citet{Santos2019a,Santos2021ApJS}, the close-in binary candidates tend to have shorter periods than the lower edge (5\textsuperscript{th} percentile) of the \prot\ distribution at a given effective temperature and tend to be beyond the upper edge (95\textsuperscript{th} percentile) of the \avsph\ distribution. Interestingly, these targets and those classified as tidally synchronized binaries by \citet{Simonian2019} occupy the same parameter space in terms of average \prot\ and \sph, tending to be outliers in comparison with solar-like stars of similar effective temperature.

Targets fainter than \kep\ magnitude $\text{Kp}=16$ were also neglected (1,774 targets), as their light curves tend to be noisier than those of bright stars. Furthermore, to ensure that one has a more complete picture of the stellar magnetic activity, we removed light curves with a time span shorter than 12 \kep\ Quarters (3,179 targets).

Removing the targets described above\footnote{The numbers of targets listed at each cut are not exclusive, as a given target may not obey multiple selection criteria (e.g., the same target may be both faint and flagged as binary candidate).}, the final target sample for the current study comprises 44,605 targets: 41,931 main-sequence stars (426 M; 13,499 K; 20,235 G; and 7,771 F) and 2,674 subgiant stars. The evolutionary stage of the targets is determined through the EEPs from {\color{blue2}Mathur et al. (in prep.)}. The \teff\ cuts adopted to split the sample into spectral types FGKM are 6000 K, 5200 K, and 3700 K, respectively. For the final sample, we have CFOP parameters for 361 targets, APOGEE parameters for 2,026 targets, LAMOST parameters for 14,493 targets, GKSPC parameters for 26,183 targets, and KSPC DR25 parameters 1,542 targets. The Hertzsprung–Russell (HR) diagram for the target sample is shown in Appendix~\ref{app:clean}.

Appendix~\ref{sec:cleansample} provides more details about the targets that were neglected in the context of the parameters we study below. While removing targets that seem to behave normally, the selection criteria above also remove a significant amount of outliers.
Despite the attempt to remove all binary candidates as described above, there is a concern related to additional potential binaries. For this reason, in Appendixes \ref{sec:cleansample} and \ref{app:regression}, we also considered the RUWE values, but we do not find any particular bias in terms of rotation or photometric magnetic activity. 

\section{Photometric magnetic activity proxy}\label{sec:sph}

The photometric magnetic AP, \sph, is the standard deviation of light curve segments of length $5\times P_\text{rot}$ as defined by \citet{Mathur2014}. \citet{Salabert2016a,Salabert2017} show that \sph\ is a proper proxy for solar and stellar magnetic activity. However, we note that \sph\ can be a lower limit of the maximum possible photometric activity level. The \sph\ values depend on the spot visibility, which is determined by the stellar inclination angle and the spot latitudinal distribution. Furthermore, the longitudinal distribution of spots also affects \sph; for example \sph\ is relatively small when spots are on opposite sides of the star. This potential underestimation of the photometric magnetic activity may be important for individual targets, particularly with small inclination angles, but its effect should not be significant in ensemble studies.

As mentioned above, the average \sph\ values, \avsph, were adopted from \citet{Santos2019a,Santos2021ApJS} for the \kep\ solar-like stars. These were corrected for the photon-shot noise following the approach by \citet{Jenkins2010_pnoise}. 
For the VIRGO/SPM data, the correction to the photon-shot noise was computed from the high-frequency (8000-8200 $\mu \text{Hz}$) noise in the power spectrum density (PSD) for the original 60-second cadence. This estimate computed from the PSD is an upper limit to the photon-shot noise in the solar observations. 

In addition to the average \sph, in this study we also computed the temporal variation of \sph\ by computing the standard deviation of the individual \sph\ values computed over $5\times P_\text{rot}$ segments. Given that \sph\ is a proxy for magnetic activity, a time series of \sph\ over the 4 yr \kep\ mission provides a brief window into the long-term, cycle-scale variability of the star, as we demonstrate using the Sun in the following section.  The standard deviation of \sph\ therefore provides a lower-limit estimate of the amplitude of long-term variability for our sample of stars.

\section{Results}\label{sec:results}

\subsection{Photometric magnetic activity of the Sun}\label{sec:ressun}

For consistency with the \kep\ bandpass, we analyzed Sun-as-a-star VIRGO g+r photometric data \citep{Basri2010}. To compute the \sph\ over segments of length $5\times P_{\text{rot}}$, we used $\text{P}_{\text{rot}\odot}=26.43\pm1.04$ days. The rotational analysis and the individual \sph\ values for solar cycles 23 and 24 are shown in Appendix~\ref{sec:appsun}.

\begin{figure}[h!]
    \centering
    \includegraphics[width=\hsize]{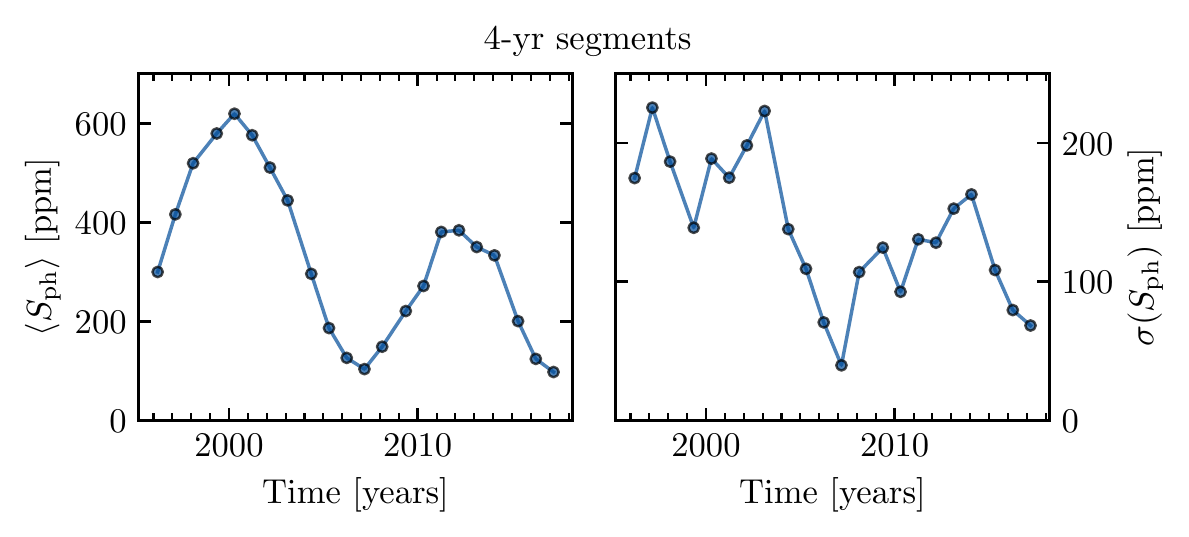}
    \caption{Average (left) and the standard deviation (right) of the individual \sph\ values over 4 yr segments for VIRGO g+r. The horizontal axis corresponds to the starting time of the 4 yr segment.}
    \label{fig:sph_sun}
\end{figure}

Next, we split the \sph\ data into segments of 4 years, which is the maximum length of the \kep\ observations. The segments are spaced by 1 year in order to cover a wide range of phases of the activity cycle. We also consider one last segment with a length of about 3 years, which is the minimum observational length considered in this work (Sect.~\ref{sec:sample}). In total, we have 22 \sph\ segments. For each segment we computed the average and the standard deviation of the individual \sph\ values (shown in Fig.~\ref{fig:sph_sun}).

The average \sph\ follows the 11 yr magnetic activity cycle of the Sun, while the \stdsph\ tends to be largest at high activity levels, but is not exactly in phase with \avsph. In fact, one can see the signature of the so-called quasi-biennial oscillations (QBOs) in the 4-year \stdsph\ around the activity maxima (see also SA and \F\ behavior in Appendix~\ref{sec:appsun}). Quasi-biennial oscillations are quasi-periodic short-term variations in the solar magnetic activity and are seen in different magnetic APs of the Sun \citep[e.g.,][]{Broomhall2012,Bazilevskaya2014,Broomhall2015,Mehta2022}. They are present at all phases of the 11 yr solar cycle, being modulated by it. However, because we are using 4 yr segments, in Fig.~\ref{fig:sph_sun}, the QBO signature is only found around the maxima, where the QBO amplitude is the largest.

\subsection{Photometric magnetic activity of the solar-like stars}\label{sec:ressph} 

In this work, we investigate the relation between the average photometric magnetic activity, \avsph, and its variation, \stdsph, over the \kep\ observations of tens of thousands of solar-like stars with detectable rotational modulation. Furthermore, we compare their \sph\ behavior with that of the Sun.

\begin{figure}
    \centering
    \includegraphics[width=\hsize]{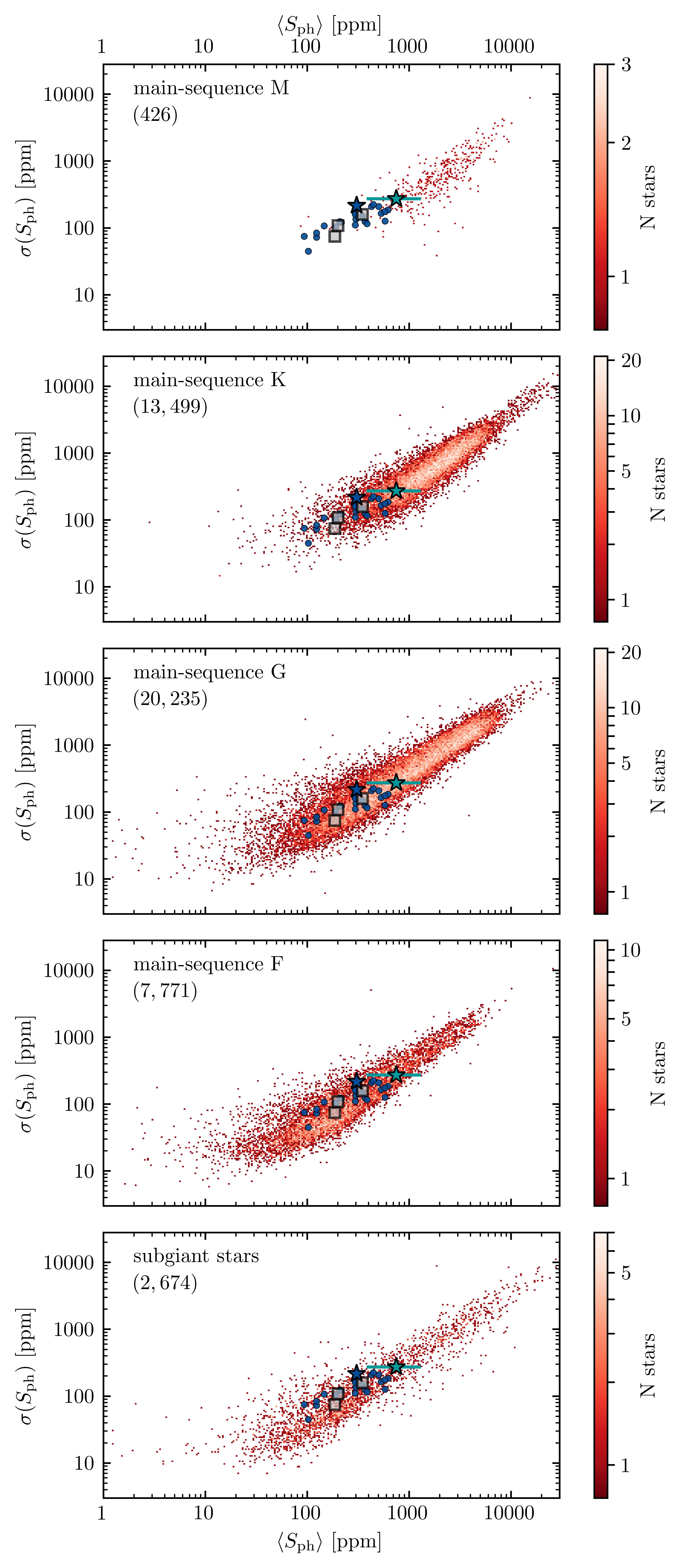}
    \caption{Standard deviation of \sph\ as a function of the average \sph\ for main-sequence stars of different spectral types and subgiants, color-coded by the number of stars. The blue star marks the solar values from the 24 yr VIRGO g+r, while the blue circles indicate the values obtained from 4 yr segments. The turquoise star highlights Doris (KIC~8006161), a solar analog with a confirmed 8 yr activity cycle. The turquoise line indicates the range between minimum and maximum individual \sph\ values. The squares mark three activity-cycle candidates: KIC~5184732, KIC~7970740, and KIC~10644253. The number of \kep\ targets of each type is indicated in each panel.}
    \label{fig:sph_std_N}
\end{figure}

Figure~\ref{fig:sph_std_N} shows the \stdsph\ versus \avsph\ relation for the main-sequence stars of different spectral types and also for the subgiant stars. We find a tight relation between \stdsph\ and \avsph, with \stdsph\ increasing with \avsph. This suggests that stars that are on average photometrically more active are also more variable in time. Indeed, given that \sph\ has been shown to be a valid magnetic AP, this result is not surprising and it is consistent with the results from ground-based spectroscopic observations, where different authors found that stars with large chromospheric emissions also exhibit large temporal variations \citep[e.g.,][]{Wilson1978,Radick1998,Radick2018,Lockwood2007,EgelandThesis, GomesDaSilva2021,EBrown2022}. Furthermore, \citet{EBrown2022} showed that the surface-averaged magnetic field strength also exhibits a similar behavior, with its temporal variation being the largest for stars with the strongest average magnetic field. 

The results for the Sun are overplotted in blue in the different panels of Fig.~\ref{fig:sph_std_N}. The blue star marks the standard deviation and average solar \sph\ values measured from full VIRGO g+r light curve ($\sim 24$ yr). The blue circles correspond to the \avsph\ and \stdsph\ measured from 4-year segments, representing different phases of the cycle (Sect.~\ref{sec:ressun}). While the 4-year \avsph\ and \stdsph\ for the Sun span a wide range of values, they tend to follow the general relation seen for the \kep\ solar-like stars, with phases of high activity level having greater variations in comparison with the low activity phases. However, there is some deviation from the general trend, for example around the solar maximum (see Fig.~\ref{fig:sph_sun}), where the signature of the QBO is present (Sect.~\ref{sec:ressun}). This translates into a larger scatter in Fig.~\ref{fig:sph_std_N} around larger values of solar \avsph, shown in blue.

KIC~8006161 (HD~173701) is also highlighted in Fig.~\ref{fig:sph_std_N} in turquoise. KIC~8006161, popularly known as Doris, is a seismic solar analog with a confirmed $\sim 8$ yr magnetic activity cycle. KIC~8006161, with a surface $P_\text{rot}$ of about 30 days, is rotating more slowly than the Sun, but it is reported to be significantly more active \citep{Karoff2018}. In addition of being a well characterized G dwarf, its activity cycle was confirmed with independent spectroscopic ground-based observations \citep{Karoff2018} spanning almost 20 years and also being partially contemporaneous to the \kep\ observations. The activity cycle is also detected in different APs, including asteroseismic ones. These attributes make KIC~8006161 a reference \kep\ solar-like star. KIC~8006161 is also intriguing and often seen as an outlier, which may be related to its high metallicity \citep[$\text{[Fe/H]}=0.34$;][]{Furlan2018} and enhanced magnetic activity. Details on the properties of KIC~8006161 are available in \citet{Karoff2018} and references therein. \kep\ observed part of the rising phase of one cycle, starting around the activity minimum \citep[e.g.,][]{Kiefer2017,Karoff2018,Santos2018}. The minimum and maximum \sph\ values are $\sim396$ and $\sim1262$ ppm, respectively. The turquoise line marks the range between these values, while the turquoise star indicates the average and standard deviation of the individual \sph\ over the 4 yr \kep\ observations. We also highlight three other stars (squares) that show evidence for cyclic magnetic activity both seen through \sph\ and asteroseismic indicators \citep[e.g.,][]{Salabert2016,Santos2018}. However, these potential magnetic cycles are not independently confirmed, and thus they should be still considered as candidates. From less to more photometrically active, the three stars are: KIC~5184732 (G~dwarf), KIC~7970740 (G~dwarf), and KIC~10644253 (F~dwarf, a young seismic solar analog). Similarly to the Sun, KIC~8006161 and the other highlighted stars follow the observed \stdsph\ versus \avsph\ relation. All four \kep\ stars are part of the asteroseismic LEGACY sample \citep{Lund2017,SilvaAguirre2017}.

In the \stdsph\ versus \avsph\ relation shown in Fig.~\ref{fig:sph_std_N}, there are underlying dependences on other stellar properties. For example, \avsph\ depends on \prot\ and \teff\ \citep[e.g.,][]{Santos2019a,Santos2021ApJS} and it may also depend on $L$, which changes as stars evolve and become less active. Thus, to make sure that we account for all potential bias, we perform a multivariate linear regression to account for dependences on different parameters: \kep\ magnitude (Kp), observational length (\tobs), effective temperature (\teff), luminosity ($L$), metallicity [Fe/H], \prot, and \avsph. Here we consider the logarithm of the \stdsph\ and \avsph.
Table~\ref{tab:residuals} lists the Spearman correlation coefficients (SCCs), which were computed between a given property and the residuals after isolating the dependence on that property (i.e., removing the dependences on the remainder). We only find significant and systematic correlations with \prot\ and log~\avsph. Appendix~\ref{app:regression} summarizes the coefficients of the multivariate regression. 

The SCC values are consistent with no or weak correlation between log~\stdsph\ and \tobs\ or Kp. The positive SCC values for Kp indicate that \stdsph\ increases toward faint stars, which typically have noisier light curves in comparison with bright stars. While there is a photon-shot noise correction applied to the \sph, its variation can still be slightly affected by noise. The SCC values are negative for \tobs, indicating that shorter light curves have smaller \stdsph, which could be expected due to the smaller number of available data points. The SCC values for \teff, $L$, and [Fe/H] are also consistent with no or weak correlation. Most of the atmospheric parameters are photometric (Sect.~\ref{sec:datasample}). Nevertheless, when considering solely the targets with spectroscopic constraints, we obtain similar results (Appendix~\ref{app:regression}). The SCC values for \prot\ are mostly consistent with moderate correlation, with the contrasting cases being the F~dwarfs and the subgiants, which show very weak correlation between \stdsph\ and \prot. This correlation with \prot\ is additional to the expected dependence in Fig.~\ref{fig:sph_std_N}: \avsph\ generally increases with decreasing rotation period. The correlation between \stdsph\ and \prot\ can also be seen if we split the data into small ranges of \sph: at roughly constant \sph, slow rotators are less variable than fast rotators. Finally, as also suggested by Fig.~\ref{fig:sph_std_N}, the correlation between log~\stdsph\ and log~\avsph\ is very strong for all main-sequence spectral types and subgiants.

In Table~\ref{tab:residuals}, we also list the 5\textsuperscript{th} and 95\textsuperscript{th} percentiles of the \avsph\ distribution. The M dwarfs in the sample present large \avsph\ and, thus, large \stdsph, while F dwarfs tend to have smaller \avsph\ and \stdsph\ in comparison with the cooler stars. The small \avsph\ values measured for F dwarfs (i.e., weak magnetic activity) may be a consequence of the shallower envelopes and less efficient dynamos unable to produce stellar winds to efficiently brake the stars. Therefore, F stars, specifically those above the Kraft break, remain fast rotators in the main sequence \citep[e.g.,][]{vanSaders2013}. Moreover, high-mass stars are also expected to have shorter activity lifetimes than the low-mass stars \citep{Reiners2012}.

\begin{figure}[h]
    \centering
    \includegraphics[width=\hsize]{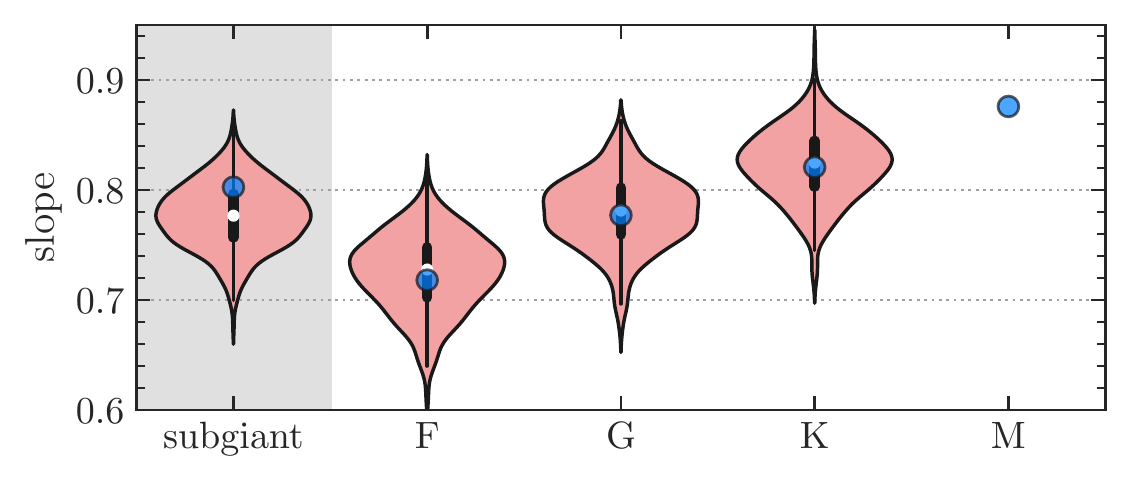}
    \caption{Resampling results for the slope of the \stdsph\ versus \avsph\ relation. The smallest subsample corresponds to the M dwarfs with only 426 targets. The resampling was done by randomly selecting 426 targets of the remaining subsamples and was repeated 500 times. For each realization, the multivariate linear regression was performed, and the final distributions for the slope are shown in light red. The thick black bars indicate the first and third quartiles, while the white circles indicate the second quartile. The blue circles indicate the values from the general multivariate regression performed with the full subsamples. The gray region marks the subgiant stars as their temperature does not follow the order of the main-sequence stars: for the dwarfs, \teff\ increases from right to left.}
    \label{fig:par_dist}
\end{figure}

The results from the multivariate regression indicate that the relation between \stdsph\ and \avsph\ becomes steeper as effective temperature decreases (from F to M). The slope values from the multivariate regression described above are listed in Appendix~\ref{app:regression} and are shown by the blue circles in Fig~\ref{fig:par_dist}. Trying to verify that there is no selection bias resulting from the different number of stars in each subsample, we conduct a resampling exercise. The smallest sample corresponds to the main-sequence stars of spectral type M (426 targets). Thus, we randomly select 426 stars from each remaining subsample (FGK dwarfs and subgiants) and we repeat the random selection 500 times. For each realization we perform a multivariate regression (as described above). The respective distributions of the slope for the log~\stdsph-log~\avsph\ relation are shown in light red in Fig.~\ref{fig:par_dist}. The results retrieved from the resampling exercise are consistent with the general results, indicating that the relation is indeed steeper for the cooler stars in comparison with the hotter stars.

\subsubsection{$P_{\rm rot}$ bimodality: Fast- and slow-rotating populations}\label{sec:branches}

As described in the introduction, the rotation-period distribution for \kep\ solar-like stars is bimodal \citep[e.g.,][]{McQuillan2014,Santos2019a,Santos2021ApJS}. While the \stdsph\ versus \avsph\ relation appears to be rather continuous, we pose the question of whether there is any substantial difference between the slow- and fast-rotating populations in terms of \sph\ variation.

In the \prot-\teff\ diagram, the two regimes are separated by a lower density region, first noticed by \citet{McQuillan2014}. The location of the period gap depends on the spectral type. In the activity-rotation diagram, one can also identify two regimes. While for the G dwarfs the fast-rotating regime seems almost saturated, it is clear for cooler stars that the fast-rotating regime is still rotation dependent. Comparably to what \citet{Reinhold2020} did for K2 stars, we split the two regimes by determining the local activity minimum between the populations in the activity-rotation diagram. The detailed procedure is described in Appendix~\ref{sec:approt}. 

Figure~\ref{fig:seq} shows the \stdsph\ versus \avsph\ relation for the fast- and slow-rotating populations for main-sequence GKM stars. The relation between the temporal variation of the photometric magnetic activity and its average values for the two populations does not seem to be distinct. For the K and G dwarfs, the stars in the fast-rotating population tend to have larger \avsph\ and, thus, larger \stdsph\ in comparison with those in the slow-rotating population. Also, for all spectral types, the median luminosity is larger for the slow-rotating population than for the fast-rotating population, which generally is consistent with the slow rotators being more evolved than the fast rotators. Using the coefficients found for the multivariate regression for the full M-, K-, and G-dwarf samples, we correct each population for the dependences on the different observational and stellar properties. We find that the distributions of the log~\stdsph\ residuals are similar and do not show a systematic trend. The most noticeable difference is that the relation (and distribution of the residuals) is tighter (narrower) for the fast-rotating population. We note, however, that the number of stars in the slow-rotating population is significantly larger than that of the fast-rotating population.  We also perform a multivariate regression separately for each population, and there are no other clear systematic differences between the two populations. Table~\ref{tab:residuals} and the tables in Appendix~\ref{app:regression} summarize the properties of the \stdsph\ versus \avsph\ relation and multivariate regression.

\begin{figure}[h]
    \centering
    \includegraphics[width=\hsize]{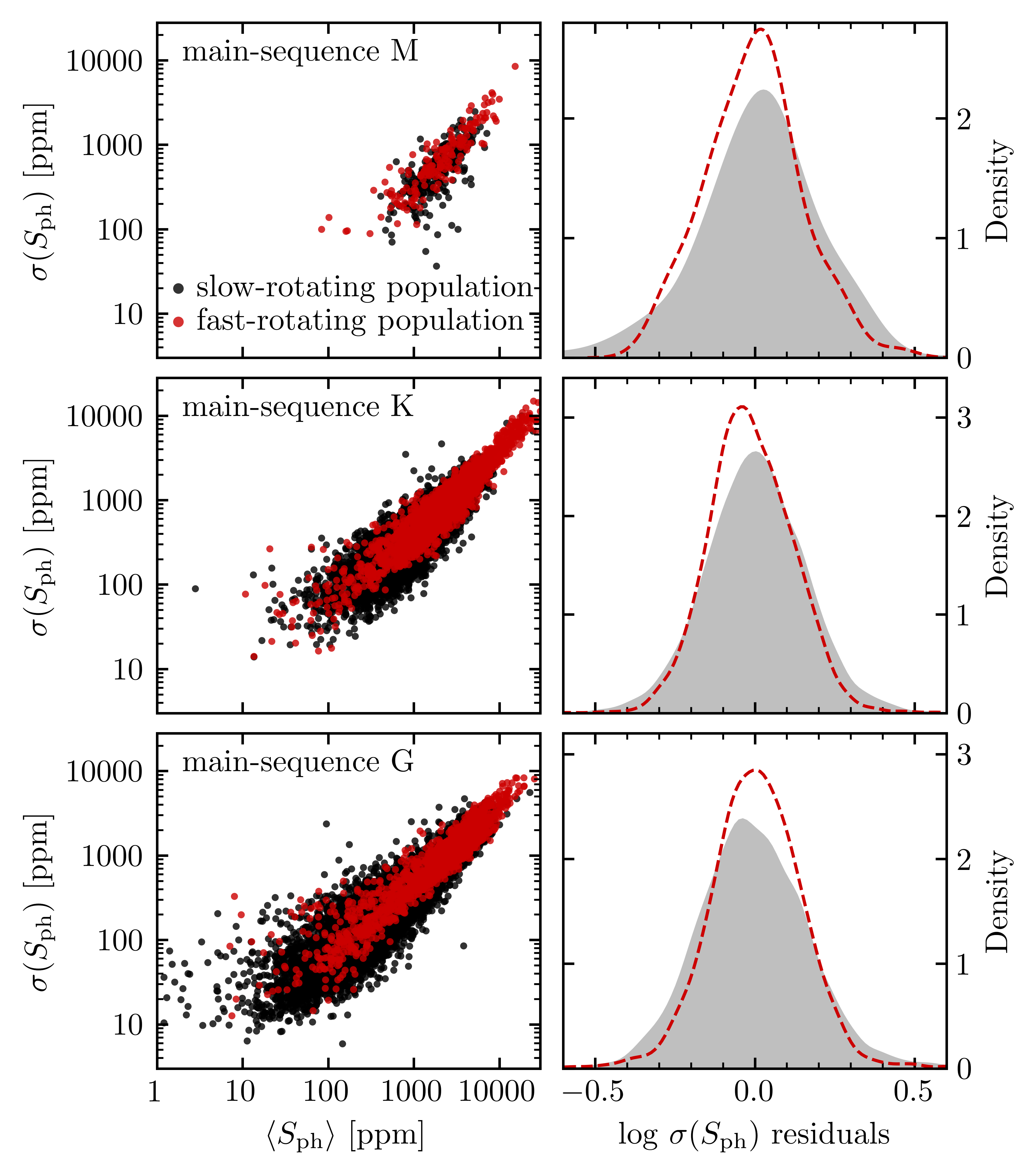}
    \caption{Relation between \stdsph\ and \avsph\ for the two populations of stars, above and below the intermediate-\prot\ gap. {\it Left}: Same as in Fig.~\ref{fig:sph_std_N} but for main-sequence GKM stars.\ The color code indicates whether they belong to the fast- (red) or slow-rotating (black) populations. For F stars ($T_\text{eff}\ge6000$ K), the rotation-period distribution is unimodal, and, thus, F stars are not considered here. {\it Right}: Distribution of the log \stdsph\ residuals computed after accounting for the dependences on the different parameters. The distributions for the fast- (red) or slow-rotating (gray) populations are illustrated by the KDE.} 
    \label{fig:seq}
\end{figure}

\subsubsection{Absence of the VP gap in Kepler photometric data}\label{sec:VP}

\begin{figure}[h!]
    \centering
    \includegraphics[width=\hsize]{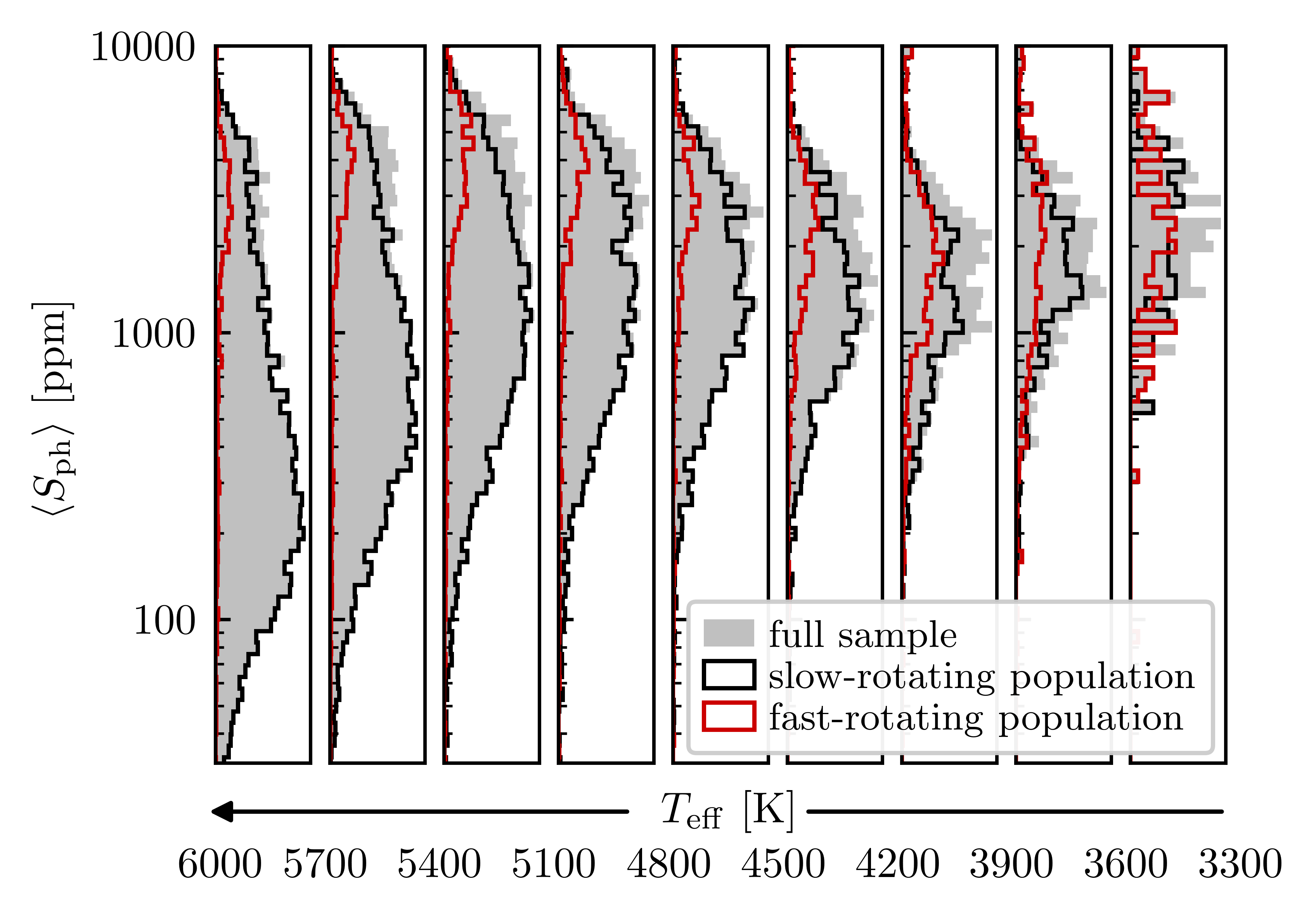}
    \caption{\avsph\ distribution in logarithmic scale for the GKM dwarfs, split into \teff\ intervals of 300 K. The gray shaded histogram shows the distribution for the full subsamples, while the solid black and red lines show the histograms for the slow- and fast-rotating populations, respectively. For clarity, the distributions here are shown by the actual histograms, while we opt for the KDE for the remainder.}
    \label{fig:Sph_branches2}
\end{figure}

In other stellar samples, there is a lack of stars with intermediate average chromospheric emission \citep[$\log\, R'_\text{HK}\sim-4.75$; e.g.,][]{VaughanPreston1980,Vaughan1980,Henry1996,GomesDaSilva2021}, which is known as VP gap. \citet{GomesDaSilva2021} and \citet{EBrown2022} recently studied the average chromospheric emission and its variability for relatively large samples of stars (1,674 and 954 stars, respectively). While \citet{GomesDaSilva2021} found gaps in the data consistent with the VP gap, \citet{EBrown2022} found under-density regions, which are less pronounced than the VP gap.

In the context of the \kep\ sample, the VP gap would happen in the slow-rotating population. In Fig.~\ref{fig:sph_std_N}, there is no evidence for gaps or discontinuities in the \avsph\ data for \kep\ stars. Figure~\ref{fig:Sph_branches2} shows the \avsph\ distribution for the GKM dwarfs in narrow \teff\ intervals. The distributions for the slow- and fast-rotating populations are also plotted individually (solid-line histograms). In this figure, instead of using the kernel density estimate (KDE), we opt to show the histograms, so it is clear that the absence of gaps is not due to any kind of smoothing. We find that the \avsph\ distribution does not show clear gaps or lower density regions. Some of the bins for M dwarfs have lower density than the neighbor bins, but it might be due to the small number of M dwarfs (Table~\ref{tab:residuals}). The \avsph\ distributions for the fast- and slow-rotating populations are the least distinct for M dwarfs (see also the values corresponding to the 5\textsuperscript{th} and 95\textsuperscript{th} percentiles in Table~\ref{tab:residuals}). For K and G dwarfs, the distributions become more distinct, with the fast-rotating population having mostly large \avsph\ values. Nevertheless, while the distributions may not be exactly unimodal, this difference does not lead to a discontinuity in the distribution. Furthermore, we recall that the VP gap would be located in the slow-rotating population, not at the intermediate-\prot\ gap. The distribution for the slow-rotating population also does not show evidence for discontinuity. 

\subsection{The Sun and the Sun-like stars}\label{sec:sunlike}

It has been a long-lasting question whether the Sun is a typical or unusual Sun-like star \citep[e.g.,][]{Soderblom1985,Lockwood1992,Henry1996,Gustafsson1998,EgelandThesis}. Recently, \citet{Reinhold2020Sci} found that stars similar to the Sun observed by \kep\ are notably more active than the Sun. This result was later questioned by \citet{Metcalfe2020_comment}, who showed that the stellar sample was biased toward lower \teff\ and higher \metal\ in comparison to the Sun. In consequence, the selected stars might have deeper convection zones in comparison to the Sun, which in turn may be behind the strong magnetic activity measurements.

In this section we compare the Sun with the Sun-like stars, which are selected according to their effective temperature, surface gravity, and rotation period. Naturally, since the study by \citet{Reinhold2020Sci}, there were updates to these stellar properties. In Appendix~\ref{sec:appsun-like} for the Sun-like stars, we compare the updated parameters with those adopted by \citet{Reinhold2020Sci} from \citet{Mathur2017}. In this work, we adopted spectroscopic parameters when available \citep{Zhao2012,Furlan2018,Zong2020,Ahumada2020}. In practice, however, most of the adopted parameters turn out to be photometric ones from \citet{Berger2020}, who took into account {\it Gaia} data and provided an update to the catalog by \citet{Mathur2017}. Furthermore, rotation periods of \kep\ solar-like (FGKM) stars have also been updated \citep{Santos2019a,Santos2021ApJS}. Particularly, one of the most significant contributions in terms of new detections of rotational modulation and the respective period corresponds to targets close to the upper edge of the rotation distribution, where the Sun lies. Finally, the photometric magnetic activity metric used here and in \citet{Reinhold2020Sci} are different. Differences between the two metrics are discussed for a sample of seismic targets with detected surface rotation in \citet{Garcia2014}. {\color{blue2}Mathur et al. (in prep.)} performs a detailed and extended comparison for the full sample of solar-like stars (not only seismic targets). The most significant discrepancies are at very low and very high activity levels.

\begin{figure}%[h]
    \centering
    \includegraphics[width=\hsize]{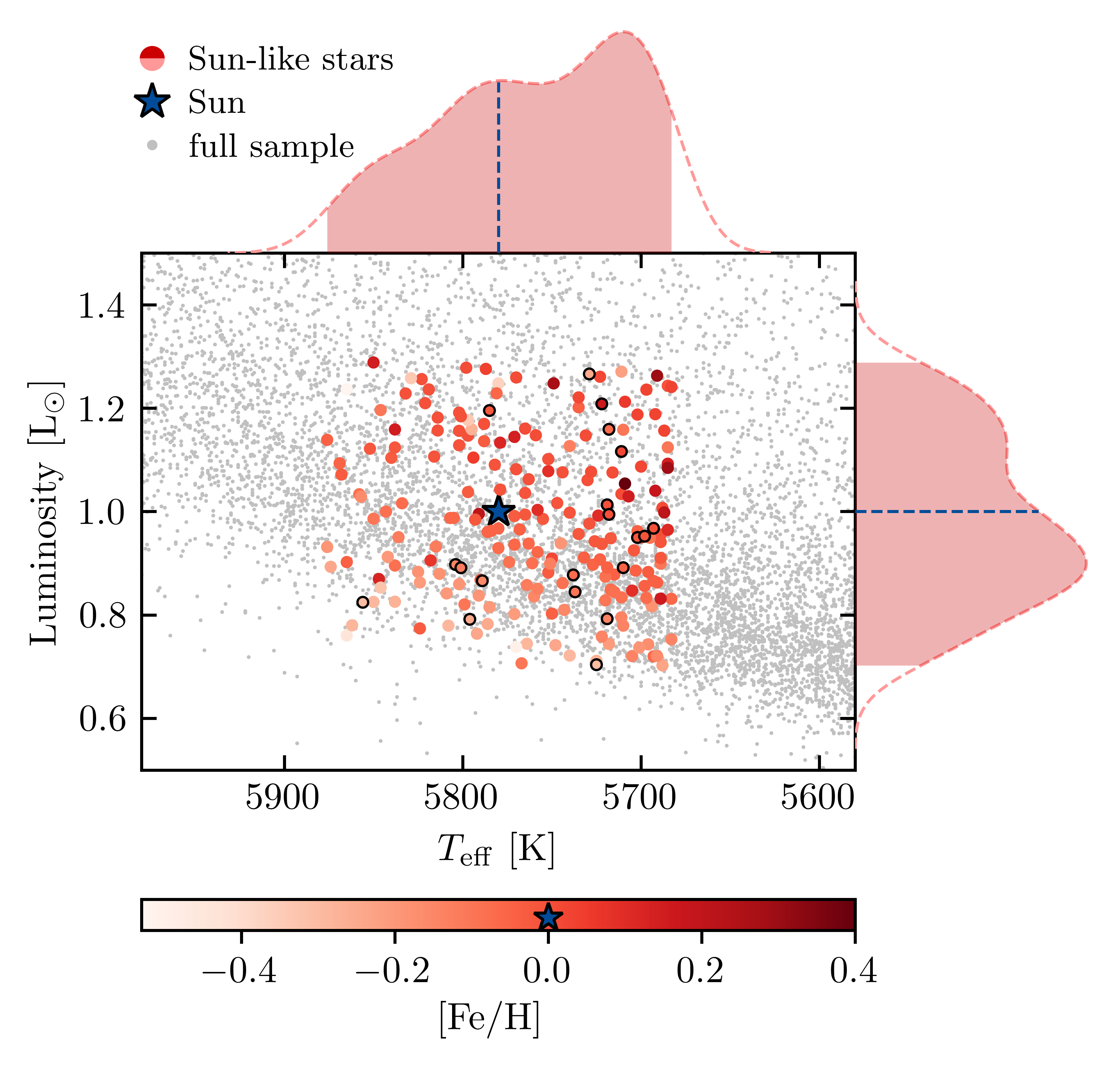}
    \caption{HR diagram, where the 211 Sun-like stars (according to their \teff, $L$, and \prot) are highlighted in shades of red, which indicate the respective metallicity. The Sun is marked by the blue star. The top and right panels show the \teff\ and $L$ distributions, where the dashed blue lines mark the solar values. The symbols with a black outline mark the targets with spectroscopic constraints. As a reference, the gray dots show targets from the full target sample.}
    \label{fig:hr_sun}
\end{figure}

As our G-dwarf subsample has a large number of targets, we can define a relatively narrow parameter space. The stars we selected as Sun-like stars have \teff\ within 100 K, $\log L$ within 0.3 (in $\text{L}_\odot$), and \prot\ within 2 days to those of the Sun. This selection results in 211 Sun-like stars. We considered the following solar values: $\text{T}_{\text{eff}\odot}=5780$ K and $\text{P}_{\text{rot}\odot}=26.43$ days. 
Figure~\ref{fig:hr_sun} shows the HR diagram highlighting the Sun-like stars in shades of red, which indicate the metallicity of the stars. Below we discuss in detail and correct for the underlying dependences or biases. 

\begin{figure}[h]
    \centering
    \includegraphics[width=\hsize]{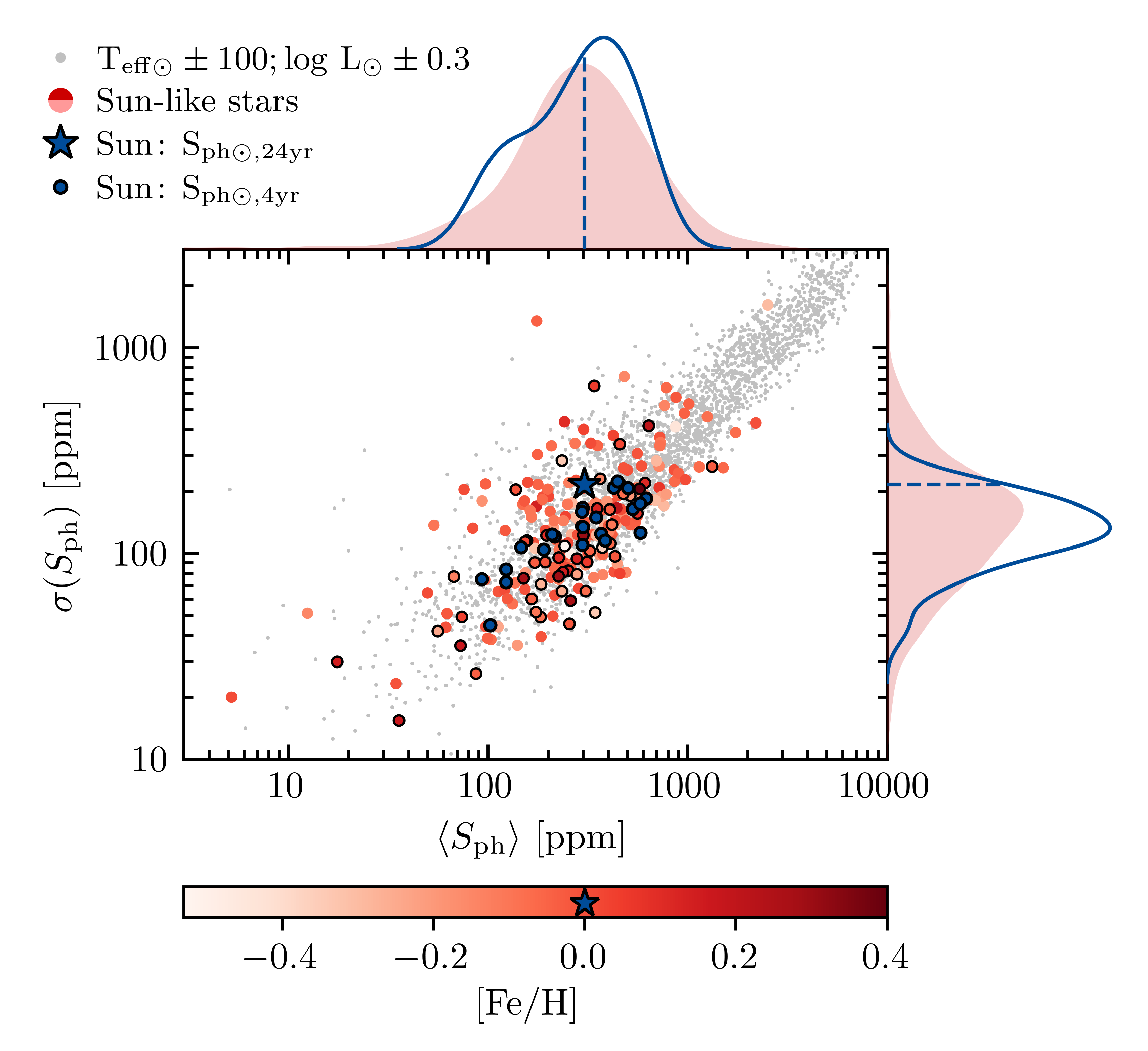}
    \caption{\stdsph\ as a function of \avsph\ for the Sun and the Sun-like stars. For reference, the gray dots show the G dwarfs with similar \teff\ and $L$, but any \prot. The Sun-like stars are indicated in red, and the shade represents the metallicity. The respective distributions are shown in red on the top and right panels. The outlined targets have spectroscopic parameters. The results from the 4 yr VIRGO g+r segments are shown with the blue circles, and the respective distributions are shown with the solid blue line. The blue star in the main panel and the dashed blue lines correspond to the full light curve, and the blue star in the color bar indicates solar metallicity.}
    \label{fig:stdsph_sunlike}
\end{figure}

Figure~\ref{fig:stdsph_sunlike} shows the same as Fig.~\ref{fig:sph_std_N} but highlighting the Sun and the Sun-like stars (shades of red). The color code indicates the metallicity of the targets and those with spectroscopic constraints are marked by the black outline. The top and right-hand panels show the respective distributions for the Sun-like stars (red) and for the 4 yr solar segments (solid blue line). The blue dashed lines mark the values computed from the full 24 yr solar light curve. For this very narrow parameter space, $\langle \text{S}_{\!\text{ph}\odot}\rangle$ and $\sigma( \text{S}_{\!\text{ph}\odot})$ do not have a distinct behavior with respect to the Sun-like stars. However, for the Sun in the last two cycles, there are fewer epochs of large $\langle \text{S}_{\!\text{ph}\odot}\rangle$ compared to \kep\ Sun-like stars. This difference is also seen in terms of the \sph\ temporal variation. 

Similarly to the previous sections, we corrected \avsph\ and \stdsph\ for dependences on different properties. To compare with the Sun, we ignored Kp and \tobs, which, as verified above, show no significant correlations. The multivariate regressions are computed for the G stars shown in gray in Fig.~\ref{fig:stdsph_sunlike}. Then, we adopted the resulting coefficients to correct the \avsph\ and \stdsph\ for the Sun-like stars and for the Sun. The results are summarized in Table~\ref{tab:sunlike} and in Appendix~\ref{sec:approt}).

As illustrated in Fig.~\ref{fig:hr_sun}, the sample of Sun-like stars is slightly biased toward low \teff\ and low $L$ in comparison to the Sun. Both lead to a larger fraction of stars with large \avsph\ and consequently \stdsph. The $L$ bias is however expected as low $L$ stars (less evolved) are the most numerous in the full sample, which can be seen for example in the density of gray data points in the background. 
We find a moderate negative correlation between the residual \avsph\ and $L$ (Table~\ref{tab:sunlike}). 
Simultaneously, the Sun-like sample has more stars rotating slightly faster than stars rotating more slowly than the Sun. This is an expected bias as the \prot\ distribution for \kep\ G dwarfs peaks around 20 days (see Appendix~\ref{sec:appsun-like}). \prot\ and \avsph\ are strongly correlated (Table~\ref{tab:sunlike}), with activity increasing toward fast rotation (negative correlation). It is important to note that the measured \prot\ depends on the (unknown) spot latitudes in a differentially rotating star, and thus the \prot\ is interpreted as an average value of the surface rotation.
In summary, the \teff, $L$, and \prot\ biases in the sample of Sun-like stars would lead to a bias toward higher activity levels than the Sun. Nevertheless, we still find the $\text{S}\!_{\text{ph}\odot}$ to be consistent with that of the selected Sun-like stars (see below). Finally, we also find a weak correlation between metallicity and \avsph\ (Table~\ref{tab:sunlike}).

\begin{figure}[h]
    \centering
    \includegraphics[width=\hsize]{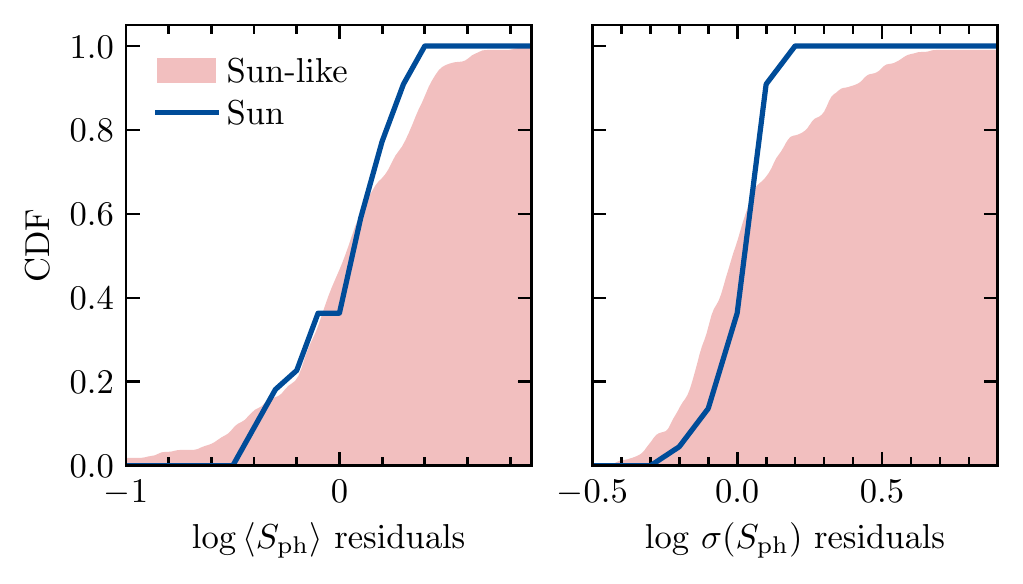}
    \caption{CDF for the log~\avsph\ and log~\stdsph\ residuals for the Sun-like stars (red shaded region) and the Sun (solid blue line).}
    \label{fig:dist_avsph}
\end{figure}

Figure~\ref{fig:dist_avsph} shows the comparison between the cumulative distribution functions (CDFs) of the residuals for the Sun (blue solid line) and the Sun-like stars (red shaded region). Concerning the average magnetic activity (left), the distribution of the $\log\,\langle \text{S}_{\text{ph}\odot}\rangle$ residuals is consistent with that of Sun-like stars. The p-value from the Kolmogorov-Smirnov test \citep{Kolmogorov1933,Smirnov1939} is 0.30, which supports the null hypothesis of the two samples ($\langle \text{S}_{\!\text{ph}\odot}\rangle$ and \avsph\ for \kep\ Sun-like stars) were drawn from the same distribution. The comparison between the two CDFs also illustrates that there are more stars with low and high activity than the Sun had epochs of such low or high activity during the past two cycles. Concerning the variation of the magnetic activity, the results for the Sun and the Sun-like stars are also consistent (p-value of 0.11). However, there is a larger fraction of highly variable stars in comparison with the variability of the $\text{S}_{\!\text{ph}\odot}$.

\subsection{Doris and the Doris-like stars}\label{sec:dorislike}

\begin{figure}[h]
    \centering
    \includegraphics[width=\hsize]{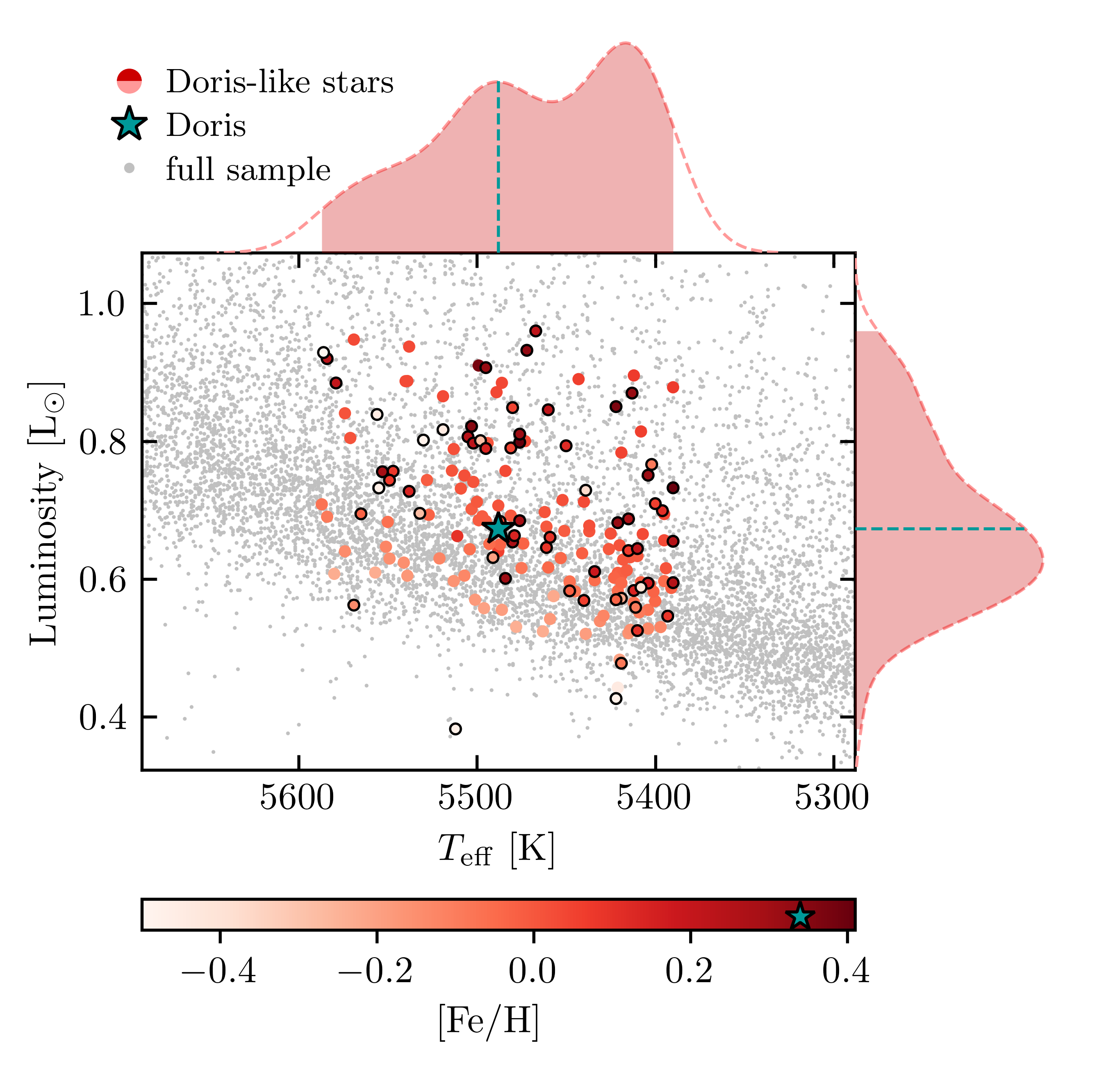}
    \caption{Same as in Fig.~\ref{fig:hr_sun} but for Doris and Doris-like stars.}
    \label{fig:hr_doris}
\end{figure}

In this section we compare Doris (KIC~8006161) with the Doris-like stars. Doris usually stands out when compared with other solar-like stars \citep[namely FGK dwarfs; e.g.,][]{Karoff2018,Santos2019b}. Its peculiar behavior has been attributed to its high metallicity, which changes the opacity and depth of the convection zone and, thus, affects magnetic activity. Above, for the stars with similar \teff\ and $L$ to the Sun, we already found evidence for a weak correlation between \metal\ and \avsph. Here, we selected the Doris-like stars by implementing the same criteria as in the previous section, but based on the stellar properties of Doris: $T_{\text{eff},\, \text{Doris}}= 5488$ K; $L_\text{Doris}=0.67\,\text{L}_\odot$; and $P_\text{rot,\,Doris}=31.71$ days \citep[][]{Berger2020,Santos2021ApJS}. These selection criteria lead to a subsample of 173 Doris-like stars. Figure~\ref{fig:hr_doris} shows the HR diagram highlighting Doris and the Doris-like stars. The Doris-like stars are more uniformly distributed in \teff\ than the Sun-like stars, but are still biased toward low $L$. Similarly to the previous section, most of the Doris-like stars are rotating slightly faster than Doris, according to their average rotation periods (Appendix~\ref{sec:appsun-like}).

\begin{figure}[h]
    \centering
    \includegraphics[width=\hsize]{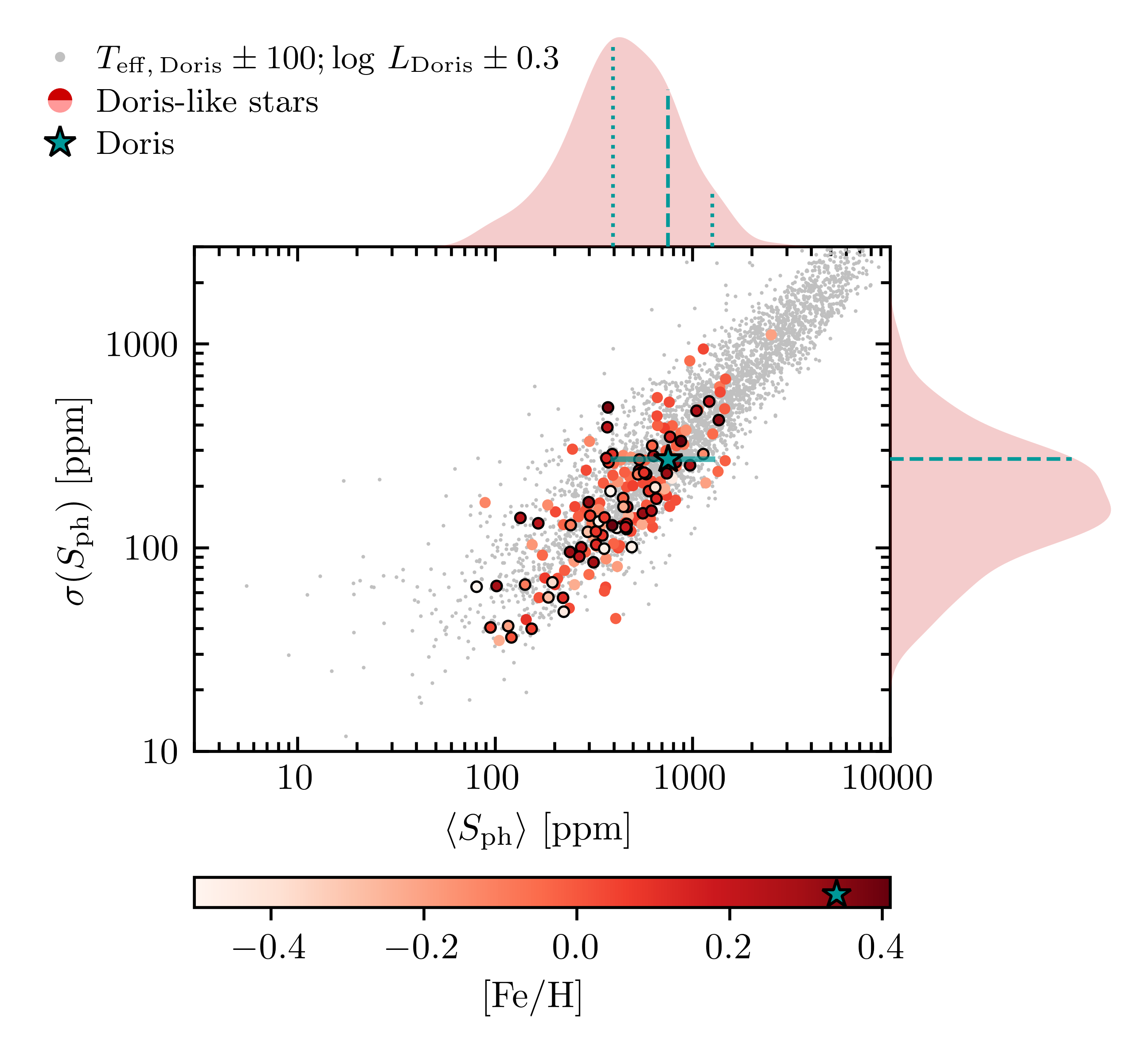}
    \caption{Same as in Fig.~\ref{fig:stdsph_sunlike} but for Doris and Doris-like stars. Doris is marked in turquoise, as in Fig.~\ref{fig:sph_std_N}. The dashed lines mark the \avsph\ and \stdsph\ for Doris, and the dotted lines mark the maximum and minimum \sph\ values (also indicated in the main panel by the horizontal line). In the color bar, the turquoise star marks the metallicity of Doris.}
    \label{fig:Dorislike}
\end{figure}

The comparison between Doris and the Doris-like stars is shown in Fig.~\ref{fig:Dorislike}. The 4-year average and standard deviation of \sph\ for Doris are marked by the turquoise star. The turquoise line indicates the minimum and maximum individual \sph\ values. The Doris-like stars are marked in shades of red, which indicate the metallicity of the targets. Interestingly, Doris-like stars seem to be more metallic than the full G dwarf sample.

When compared with stars with very similar properties, Doris is not an outlier. Nevertheless, Doris is still among the most active and most variable stars: 79\% of the Doris-like stars have a \avsph\ value smaller than $\langle S_\text{\!ph} \rangle_\text{Doris}$; and 76\% of the Doris-like stars have a \stdsph\ value smaller than $\sigma(S_\text{\!ph} )_\text{Doris}$. The lowest \sph\ value observed by \kep\ is still consistent with high activity. Indeed \kep\ observations started at the minimum of activity between two cycles \citep{Karoff2018}.  

We perform the multivariate regression based on all targets in gray (similar \teff\ and $L$ to those of Doris) and compute the correlation coefficients for the isolated dependences (Table~\ref{tab:dorislike}). Particularly, the results are consistent with a weak correlation between the \avsph\ and \metal. The evidence for a correlation between \avsph\ and \metal\ (here and in the previous section) supports the interpretation by \citet[][see also \citealp{See2021}]{Karoff2018} for Doris's strong magnetic activity.

\subsection{Relation between \prot\ and metallicity}\label{sec:metal}

Focusing solely on spectroscopic APOGEE and LAMOST metallicities for the \kep\ stars, \citet{Amard2020b} found that metal-rich stars are systematically slower rotators \citep[from][]{McQuillan2014} than their metal-poor counterparts. 

The first evidence in our sample for such a relation between \prot\ and \metal\ is found among the Doris-like stars, which were selected according to their \teff, $L$, and \prot. In the resulting sample, there is an excess of metal-rich stars in comparison to the full G-dwarf sample (see Appendix~\ref{sec:appsun-like}). This suggests that metallicity may play an important role in generating such slow rotators ($P_\text{rot}\sim31$ days) in this region of the HR diagram. 

\begin{figure}
    \centering
    \includegraphics[width=\hsize]{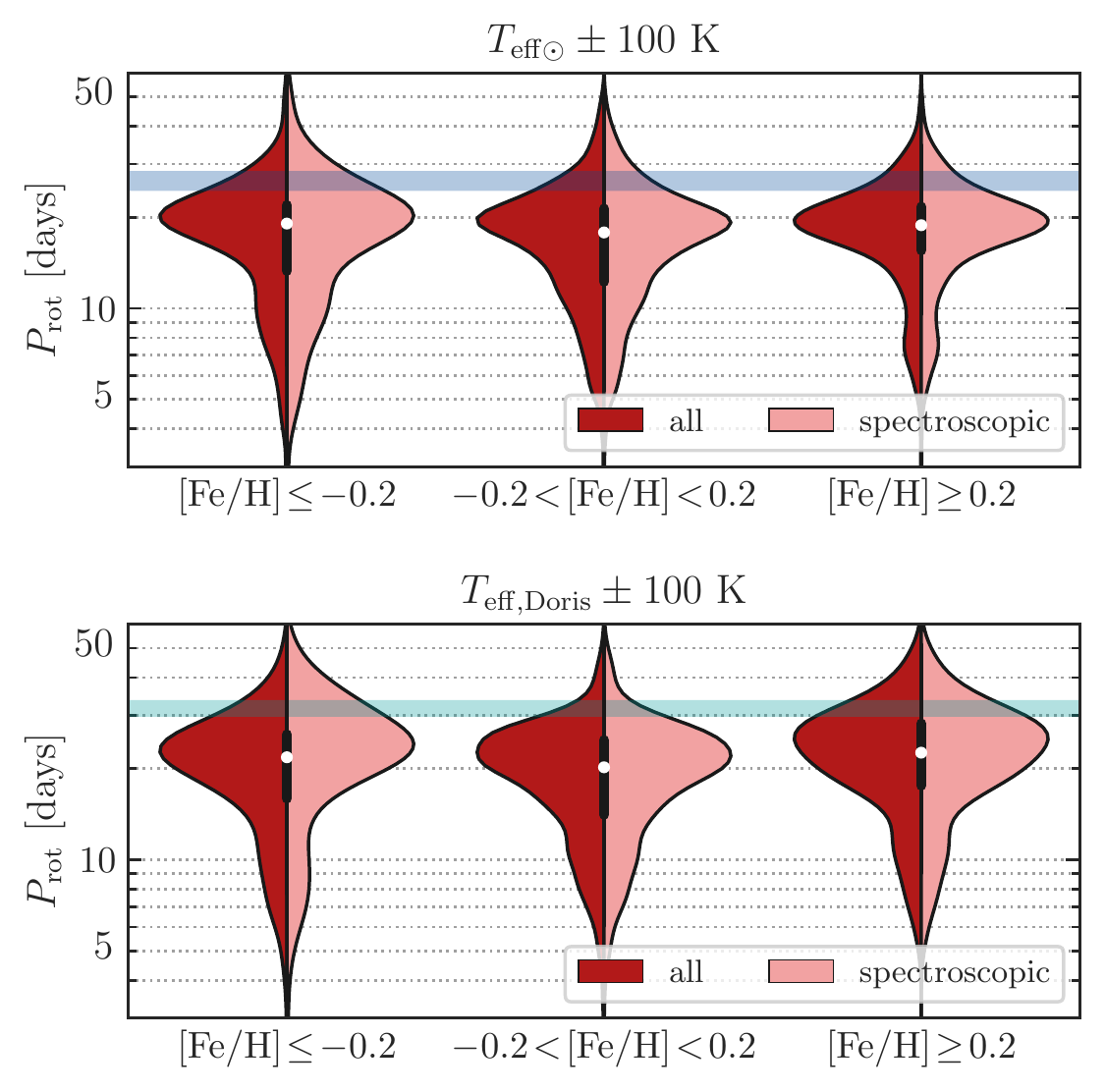}
    \caption{\prot\ distributions for metal poor (left), solar metallicity (middle), and metal rich (right) stars within 100K to the \teff\ of the Sun (top) and Doris (bottom). Dark red shows the distribution for all stars, while light red corresponds to those with spectroscopic parameters. We only consider main-sequence stars. The shaded regions in blue and turquoise mark the \prot\ interval used to select Sun-like and Doris-like stars. The white circles indicate the median values, while the black bars indicate the 1\textsuperscript{st} and 3\textsuperscript{rd} quartiles of the distributions.}
    \label{fig:metal_sun_doris}
\end{figure}

Figure~\ref{fig:metal_sun_doris} shows the \prot\ distributions for the Sun-like and Doris-like stars split according to their metallicity. We consider three \metal\ intervals: metal poor $\text{[Fe/H]}\leq-0.2$; solar metallicity $-0.2<\text{[Fe/H]}<0.2$; and metal rich $\text{[Fe/H]}\geq0.2$. The blue and turquoise shaded regions mark the interval of Sun-like and Doris-like \prot, respectively. In the parameter space of the Sun-like stars, the \prot\ distributions are similar for all \metal\ intervals. In the Doris-like \teff\ regime, the \prot\ distribution shifts toward longer \prot\ for the metal-rich subsample in comparison to the metal-poor and solar-metallicity ones.

A detailed investigation of the relationship between \prot\ and metallicity is beyond the scope of the current study. The impact of metallicity in the magnetic-activity and rotational evolution will be the focus of a future work. Nevertheless, the goal of this section was to verify the source of the excess of metal-rich stars in the Doris-like sample. Particularly, in the context of Doris being a metal-rich star, whose metal content may be behind its strong activity. Above, we confirm that, in the Doris-like \teff, the metal-rich stars are indeed slower rotators than the remainder. These results support the dependence of \prot\ on \metal\ found by \citet{Amard2020b}.

\begin{table}\small
    \centering
    \begin{tabular}{rr|cccc|c}
    \hline\hline
    & & \multicolumn{4}{c|}{main sequence} & \\
    & & M & K & G & F & subgiant \\\hline

    \multicolumn{2}{r|}{N stars} & 426 & 13,499 & 20,235 & 7,771 & 2,674\\
    \multicolumn{2}{r|}{$\langle S_\text{\!ph}\rangle_{5^\text{th}}$} & 567.1 & 281.7 & 81.4 & 30.9 & 37.4\\
    \multicolumn{2}{r|}{$\langle S_\text{\!ph}\rangle_{95^\text{th}}$} & 5950.6 & 5125.5 & 4870.7 & 2031.6 & 4239.8\\\hline
    
    \multirow{7}{*}{\rotatebox[origin=c]{90}{SCC}} & Kp & 0.07 & 0.07 & 0.15 & 0.28 & 0.21 \\
    & \tobs & 0.00 & -0.09 & -0.12 & -0.20 & -0.17 \\
    & \teff & 0.11 & 0.10 & -0.07 & -0.14 & -0.24 \\
    & $L$ & -0.11 & 0.00 & -0.11 & -0.15 & 0.04 \\
    & $\text{[Fe/H]}$ & 0.03 & 0.04 & 0.01 & -0.06 & -0.02 \\
    & \prot & -0.38 & -0.36 & -0.37 & -0.15 & -0.16 \\
    & \avsph & 0.84 & 0.91 & 0.94 & 0.88 & 0.88 \\\hline
%    \multicolumn{2}{c|}{median} & \multirow{2}{*}{0.010} & \multirow{2}{*}{-0.003} & \multirow{2}{*}{-0.008} & \multirow{2}{*}{-0.024} & \multirow{2}{*}{-0.015} \\
%    \multicolumn{2}{c|}{log \stdsph\ res.}& & & & & \\\hline
   
    \multicolumn{7}{r}{}\\
    \multicolumn{7}{l}{\textsc{Fast-Rotating population}}\\\hline
    \multicolumn{2}{r|}{N stars} & 175 & 2,848 & 2,104 & \multicolumn{2}{r}{} \\
    \multicolumn{2}{r|}{$\langle S_\text{\!ph}\rangle_{5^\text{th}}$} & 507.8 & 378.8 & 138.7 & \multicolumn{2}{r}{}\\
    \multicolumn{2}{r|}{$\langle S_\text{\!ph}\rangle_{95^\text{th}}$} & 7624.7 & 9666.7 & 6800.0 & \multicolumn{2}{r}{}\\\hline
    \multirow{7}{*}{\rotatebox[origin=c]{90}{SCC}} & Kp & 0.06 & 0.06 & 0.02 & \multicolumn{2}{r}{} \\
    & \tobs & -0.05 & -0.15 & -0.15 & \multicolumn{2}{r}{} \\
    & \teff & 0.16 & -0.03 & -0.07 & \multicolumn{2}{r}{} \\
    & $L$ & -0.11 & 0.01 & -0.07 & \multicolumn{2}{r}{} \\
    & $\text{[Fe/H]}$ & 0.12 & 0.07 & 0.01 & \multicolumn{2}{r}{} \\
    & \prot & -0.46 & -0.41 & -0.19 & \multicolumn{2}{r}{} \\
    & \sph & 0.91 & 0.93 & 0.93 & \multicolumn{2}{r}{} \\\hline

    \multicolumn{7}{r}{}\\
    \multicolumn{7}{l}{\textsc{Slow-Rotating population}}\\\hline
    \multicolumn{2}{r|}{N stars} & 251 & 10,651 & 18,131 & \multicolumn{2}{r}{}\\
    \multicolumn{2}{r|}{$\langle S_\text{\!ph}\rangle_{5^\text{th}}$} & 742.5 & 269.1 & 78.1 & \multicolumn{2}{r}{}\\
    \multicolumn{2}{r|}{$\langle S_\text{\!ph}\rangle_{95^\text{th}}$} & 4518.0 & 4465.6 & 4410.3 & \multicolumn{2}{r}{}\\\hline
    \multirow{7}{*}{\rotatebox[origin=c]{90}{SCC}} & Kp & 0.08 & 0.08 & 0.16 & \multicolumn{2}{r}{} \\
    & \tobs & 0.01 & -0.08 & -0.12 & \multicolumn{2}{r}{} \\
    & \teff & 0.18 & 0.07 & -0.06 & \multicolumn{2}{r}{} \\
    & $L$ & -0.26 & 0.00 & -0.12 & \multicolumn{2}{r}{} \\
    & $\text{[Fe/H]}$ & 0.03 & 0.04 & 0.02 & \multicolumn{2}{r}{} \\
    & \prot & -0.18 & -0.35 & -0.33 & \multicolumn{2}{r}{} \\
    & \avsph & 0.77 & 0.89 & 0.93 & \multicolumn{2}{r}{} \\\hline\hline
    \end{tabular}
    \caption{Spearman correlation coefficients (SCCs) between the log~\stdsph\ residuals, when isolating each dependence, and the different observational and stellar properties. The first part of the table concerns all the stars in Fig.~\ref{fig:sph_std_N}, which are split into slow- and fast-rotating populations (GKM) for the second and third parts of the table (Fig.~\ref{fig:seq}). The top lines of each part indicate the number of stars and 5\textsuperscript{th} and 95\textsuperscript{th} percentiles of the \avsph\ distribution.}
    \label{tab:residuals}
\end{table}\normalsize

\begin{table}\small
    \centering
    \begin{tabular}{rr|cc|cc}
    \multicolumn{6}{c}{$\text{T}_{\text{eff}\odot}\pm 100$ K; $\text{L}_\odot\pm 0.3$}\\\hline\hline
    & & \multicolumn{2}{c|}{all} & \multicolumn{2}{c}{spectroscopic}\\\hline
    
    \multicolumn{2}{r|}{N stars} & \multicolumn{2}{c|}{3470}& \multicolumn{2}{c}{1303}\\
    \multicolumn{2}{r|}{$\langle S_\text{\!ph}\rangle_{5^\text{th}}$} & \multicolumn{2}{c|}{87.0} & \multicolumn{2}{c}{72.0}\\
    \multicolumn{2}{r|}{$\langle S_\text{\!ph}\rangle_{95^\text{th}}$} & \multicolumn{2}{c|}{4454.6} & \multicolumn{2}{c}{4552.9}\\
    \\\hline
    
    & & \avsph & \stdsph & \avsph & \stdsph\\\hline
    \multirow{5}{*}{\rotatebox[origin=c]{90}{SCC}} & \teff & 0.03 & -0.02 & -0.04 & -0.03\\
    & $L$ & -0.35 & -0.10 & -0.28 & -0.13\\
    & \metal & 0.20 & 0.02 & 0.34 & 0.11\\
    & \prot & -0.67 & -0.37 & -0.75 & -0.38\\
    & \avsph & -- & 0.92 & -- & 0.94 \\\hline\hline
    \end{tabular}
    \caption{SCC between the log~\avsph\ and log~\stdsph\ residuals and the different stellar properties of stars with similar \teff\ and $L$ to those of the Sun (gray dots in Fig.~\ref{fig:dist_avsph}). The first columns correspond to all stars within the parameter space, while the last ones correspond solely to the stars with spectroscopy. The top lines indicate the number of stars and 5\textsuperscript{th} and 95\textsuperscript{th} percentiles of the \avsph\ distribution.}
    \label{tab:sunlike}
\end{table}\normalsize

\begin{table}\small
    \centering
    \begin{tabular}{rr|cc|cc}
    \multicolumn{6}{c}{$T_\text{eff,Doris}\pm 100$ K; $L_\text{Doris}\pm 0.3$}\\\hline\hline
    & & \multicolumn{2}{c|}{all} & \multicolumn{2}{c}{spectroscopic}\\\hline
    
    \multicolumn{2}{r|}{N stars} & \multicolumn{2}{c|}{4240} & \multicolumn{2}{c}{1443}\\
    \multicolumn{2}{r|}{$\langle S_\text{\!ph}\rangle_{5^\text{th}}$} & \multicolumn{2}{c|}{140.9} & \multicolumn{2}{c}{121.0}\\
    \multicolumn{2}{r|}{$\langle S_\text{\!ph}\rangle_{95^\text{th}}$} & \multicolumn{2}{c|}{5208.2} & \multicolumn{2}{c}{5496.2}\\
    \multicolumn{2}{l}{}\\\hline
    
    & & \avsph & \stdsph & \avsph & \stdsph\\\hline
    \multirow{5}{*}{\rotatebox[origin=c]{90}{SCC}} & \teff & 0.02 & -0.02 & 0.02 & -0.01\\
    & $L$ & -0.29 & -0.06 & 0.28 & -0.08 \\
    & \metal & 0.20 & 0.00 & 0.30 & 0.07\\
    & \prot & -0.61 & -0.39 & -0.65 & -0.37\\
    & \avsph & -- & 0.93 & -- & 0.95 \\\hline\hline
    \end{tabular}
    \caption{Same as Table~\ref{tab:sunlike} but for stars with \teff\ and $L$ similar to those of Doris (gray dots in Fig.~\ref{fig:Dorislike}).}
    \label{tab:dorislike}
\end{table}\normalsize

%\pagebreak

\section{Discussion and conclusions}\label{sec:conclusion}

With the advent of planet-hunting missions such as \kep, high-precision photometric time series are available for an extraordinarily large number of solar-like stars. For time series with rotational modulation, the photometric magnetic activity can be constrained through the \sph\, \citep{Mathur2014}.
In this work we investigate the temporal variability of the \sph\ over the length of the \kep\ observations. Our target sample includes more than  44,000 \kep\ solar-like (FGKM) stars with known average surface rotation periods and average \sph\ \citep[selected from][]{Santos2019a,Santos2021ApJS}. Furthermore, we analyzed solar VIRGO/SPM data in order to compare the behavior of the solar \sph\ with that observed for other stars.

We find that the photometric magnetic activity is more variable for stars that are on average more active in comparison to that of weakly active targets. This kind of relationship is known for the chromospheric magnetic activity  \citep[e.g.,][]{Wilson1978,Radick1998,Lockwood2007, EgelandThesis, GomesDaSilva2021,EBrown2022} and surface-averaged magnetic field strength \citep{EBrown2022}, but in this work we show that it is also observed in the \kep\ photometric data of solar-like stars. In particular, the standard deviation of the individual measurements of photometric magnetic activity, \stdsph, and their average value, \avsph, are very strongly correlated, independent of the spectral type. This includes the low-activity fast-rotating F stars. Due to their thin convective zones, for most of the main sequence, F stars have longer spin-down timescales in comparison to the lower-mass stars \citep{vanSaders2013,Matt2015} and, thus, do not spin down significantly, remaining fast rotators. Nevertheless, for F stars, there is still a very strong correlation between their average magnetic activity and its variation. We also find such a strong correlation for the subgiant stars that already evolved off the main sequence.

Given that the definition of \sph, \stdsph\ is a measurement of the magnetic activity variability at timescales longer than \prot. In particular, as seen for the solar data, $\sigma(\text{S}_{\text{ph}\odot})$ shows the signature of the long-term 11 yr cycle as well as the signature of the QBOs (short-term variations) at high-activity epochs. From the long-term ground-based spectroscopic observations, several other solar-like stars are known to have variations analogous to the solar QBOs as well as to the long-term activity cycle \citep[e.g.,][]{Baliunas1995,Metcalfe2013,Egeland2015}. Thus, it is possible that the 4 yr \stdsph\ for \kep\ stars also include the signature of both short-term and long-term variations in the magnetic activity. 

In addition to the very strong correlation between \stdsph\ and \avsph, our results suggest an additional dependence of the \stdsph\ on the rotation period, even after accounting for the dependence on \avsph. Although the results are only consistent with a weak correlation, at fixed \avsph\, faster rotators tend to exhibit more magnetic activity variability than slow rotators. 

GKM dwarfs in the \kep\ field fall into two populations, separated by the so-called intermediate-\prot\ gap, which is especially prominent for K and M dwarfs. For G stars, the \prot\ distribution is still bimodal, but there is no clear gap (Appendix~\ref{sec:approt}). While the fast-rotating population is on average more active than the slow-rotating one, we find that the distributions of the \stdsph\ residuals (computed using a multivariate regression) are similar for both populations. These results suggest that the relation between the mean value of \sph\ and its variability does not change. 

The Sun also follows the same \stdsph\ versus \avsph\ relationship as the other solar-like stars, both for the full 24 yr time series and for the multiple 4 yr segments. The measured values for the solar 4 yr segments illustrate, however, the limitations associated with the observation length. In particular, the 4 yr data points span a relatively wide range of values in both \stdsph\ and \avsph, depending on the phase of the cycle. Four years of \kep\ continuum monitoring might still be too short in the context of fully characterizing magnetic activity. This is especially important for the slow rotators, whose magnetic activity is expected to vary over longer timescales in comparison to the fast rotators \citep[e.g.,][]{Brandenburg1998,Bohm-Vitense2007}. Moreover, some scatter in the \stdsph\ versus \avsph\ can also be introduced by the impact of stellar inclination and spot or active-region latitudes on the amplitude of the rotational signal. In addition to the Sun, we also highlight four other stars: three with cycle candidates \citep[KIC~5184732, KIC 7970740, and KIC 10644253;][]{Salabert2016,Santos2018} and KIC~8006161 (HD~173701), also known as Doris. Doris is a well-characterized seismic G dwarf with a relatively strong magnetic activity and a confirmed activity cycle of $\sim8$ years \citep{Karoff2018}. All these solar-like stars behave in accordance to the general trend. 

While the Sun is found to be among the less active and less variable G~dwarfs observed by \kep, when compared with Sun-like stars, selected from a very narrow parameter space around the solar properties, the Sun is rather normal. This conclusion is in contrast with the findings by \citet{Reinhold2020Sci}, who find that \kep\ Sun-like stars are distinctively more active than the Sun. We must note, however, that the sample of Sun-like stars with known rotation periods has been extended since then \citep{Santos2021ApJS}. Moreover, the stellar properties adopted to select the Sun-like stars were also updated. As a result, the stellar classification has slightly changed (Appendix~\ref{sec:appsun-like}). Finally, \citet{Metcalfe2020_comment} already noticed possible selection biases.

Similarly to the Sun versus Sun-like stars comparison, we compare the well-characterized G dwarf Doris with Doris-like stars. It has been suggested that its strong magnetic activity is related to its high metal content \citep{Karoff2018}. Our results show that Doris behaves consistently with stars of very similar properties. Nevertheless, Doris is indeed among the most active stars, even at its minimum activity. Interestingly, we find an excess of metal-rich stars among the sample of Doris-like stars (Doris also being a metal-rich star). From the adopted selection criteria, the parameter that could lead to such a peculiar feature in the \metal\ distribution is \prot. Doris is a slow rotator with $P_\text{rot}\sim31.71$ days. Indeed we find that, at $\sim T_\text{eff,Doris}$, the \prot\ distribution shifts toward longer values from the metal-poor subsample to the metal-rich subsample. These results are in agreement with the findings by \citet{Amard2020b}, who also find a \prot-\metal\ dependence. Furthermore, although Doris is not an activity outlier in comparison to Doris-like stars\footnote{Doris is found to be an activity outlier in comparison to stars within a broader parameter space than that adopted in this study \citep[e.g.,][]{Kiefer2017,Karoff2018,Santos2019b}.}, we find that \metal\ still plays a role: Doris-like stars tend to be more metallic; and we find a weak correlation between \metal\ and \avsph. In fact, our results are consistent with the interpretation proposed in \citet{Karoff2018}. \citet{Karoff2018} suggested that Doris's strong magnetic activity is related to a deeper convection zone due to the increased opacity resulting from its high metal content \citep[e.g.,][]{Schwarzschild1906,vanSaders2012,Amard2020} in comparison to a metal-poor counterpart. If the magnetic activity is stronger, then the magnetic braking and the spin-down will be more efficient, leading to longer rotation periods for a given stellar age. These results are also in good agreement with the theoretical predictions by \citet{Amard2020}, who find a significant \metal\ impact on the stellar structure, consistent with metal-rich stars being more magnetically active and spinning down faster than metal-poor stars. In that way, it is important to consider chemical composition in gyrochronology studies \citep{Amard2020,Claytor2020}.

Finally, we note that the \kep\ \sph\ data do not show a discontinuity or evidence for a gap in activity that would be consistent with the VP gap, which would be located in the slow-rotating population. However, we consider that \kep\ \sph\ data cannot reject the existence of the VP gap, still debated in the literature. Firstly, while photometric and chromospheric APs are well correlated \citep{Salabert2016a,Salabert2017}, they are sensitive to different activity-related phenomena and to different layers of the stellar atmosphere. Moreover, as shown for the Sun, \stdsph\ and \avsph\ values depend on the phase of the cycle. This indicates that we can expect some scatter in the \stdsph\ versus \avsph\ relation due to the limited 4 yr \kep\ observations. This effect, together with the fact that the \kep\ sample is many times larger than those in chromospheric activity studies, can smear the existence of a lower density region or gap. Thus, our results do not exclude the existence of the VP gap. 

\begin{acknowledgements}
The authors thank Vardan Zh. Adibekyan for the helpful discussions. This work was supported by Funda\c{c}\~ao para a Ci\^encia e a Tecnologia (FCT) through national funds and by Fundo Europeu de Desenvolvimento Regional (FEDER) through COMPETE2020 - Programa Operacional Competitividade e Internacionaliza\c{c}\~ao by these grants: UIDB/04434/2020 \& UIDP/04434/2020.
This work was also supported by the National Aeronautics and Space Administration (NASA) under Grant No. NNX17AF27G to the Space Science Institute (Boulder, CO USA). This paper includes data collected by the \kep\, mission and obtained from the MAST data archive at the Space Telescope Science Institute (STScI). Funding for the \kep\ mission is provided by the NASA Science Mission Directorate. STScI is operated by the Association of Universities for Research in Astronomy, Inc., under NASA contract NAS 5–26555. ARGS acknowledges the support from the FCT through the work contract No. 2020.02480.CEECIND/CP1631/CT0001. SM acknowledges support from the Spanish Ministry of Science and Innovation (MICINN) with the Ram\'on y Cajal fellowship no.~RYC-2015-17697, PID2019-107061GB-C66 for PLATO, and through AEI under the Severo Ochoa Centres of Excellence Programme 2020--2023 (CEX2019-000920-S). SM, AJ, and DGR acknowledge support from the Spanish Ministry of Science and Innovation (MICINN) grant no.~PID2019-107187GB-I00. RAG, SNB, and LA acknowledge the support from PLATO and GOLF CNES grants. A-MB acknowledges the support from STFC consolidated grant ST/T000252/1. TSM acknowledges support from NASA grant 80NSSC22K0475. MSC acknowledges the support from the FCT through a work contract (CEECIND/02619/2017).\\

\textit{Software:} AstroPy \citep{astropy:2013,astropy:2018}, KADACS \citep{Garcia2011}, ROOSTER \citep{Breton2021}, \texttt{kiauhoku} \citep{Claytor2020,Claytor2020b}, Matplotlib \citep{matplotlib}, NumPy \citep{numpy}, Scikit-learn \citep{scikit-learn}, SciPy \citep{scipy}, Seaborn \citep{seaborn}, pandas \citep{mckinney-proc-scipy-2010_pandas,reback2020pandas}

\end{acknowledgements}

%\setstretch{-0.1}
\bibliographystyle{aa}
\bibliography{Sph_variation,python,Gaia}

\appendix

\section{Solar rotation and magnetic activity}\label{sec:appsun}

\begin{figure}[h]
    \centering
    \includegraphics[width=\hsize]{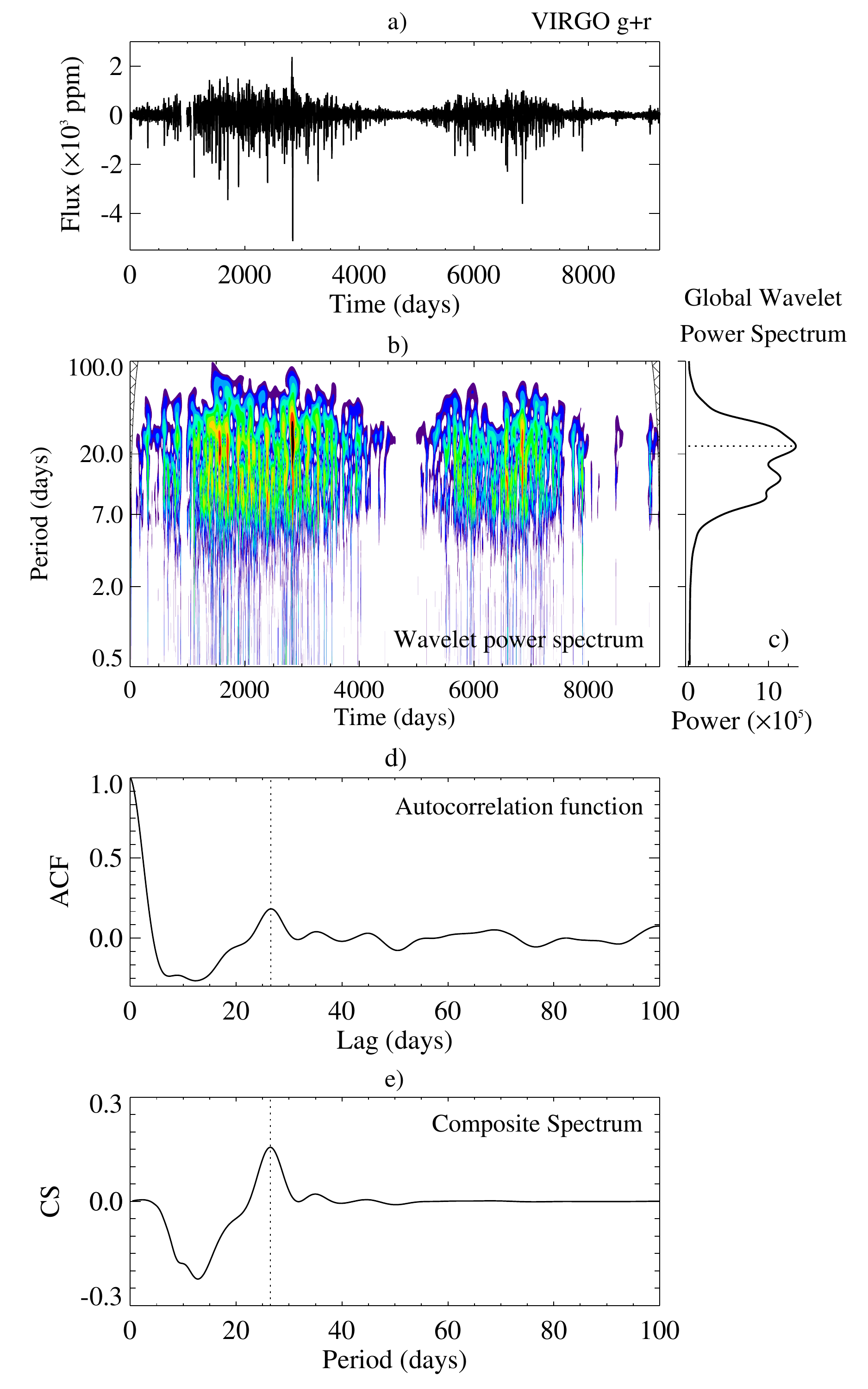}\vspace{-.2cm}
    \caption{Rotational analysis for the Sun. a) Solar VIRGO g+r light curve for more than 24 years, accounting for solar cycles 23 and 24. b) Wavelet power spectrum (WPS), where we adopt the Morlet wavelet. The color code corresponds to Rainbow+White, where black and white correspond to the highest and the lowest power, respectively. c) GWPS, which is the sum of power along the x-axis of the WPS (i.e., time). d) ACF of the light curve. e) CS, which is the product between the normalized versions of c) and d). The dotted lines indicate the three period estimates.}
    \label{fig:rot_VIRGO}\vspace{-.1cm}
\end{figure}

In order to compare the Sun with other solar-like stars, we use Sun-as-a-star VIRGO g+r (Sects.~\ref{sec:data} and \ref{sec:ressun}). Here, we summarize the results from the rotational analysis of VIRGO g+r. Figure~\ref{fig:rot_VIRGO} shows the VIRGO g+r light curve and the results from the rotation diagnostics employed in our rotation pipeline \citep[e.g.,][]{Mathur2010b,Garcia2014,Ceillier2016,Ceillier2017}.\vspace{-.2cm}

The rotation pipeline employs wavelet analysis, the ACF, and the CS. The latter is indeed the composition between the former two, corresponding to the product of their normalized counterparts. We favor the \prot\ estimate from the global wavelet power spectrum (GWPS) because of the conservative uncertainty, which for \kep\ solar-like stars is on average 10\% of \prot\ \citep{Santos2021ApJS}. The ACF and CS serve mostly as validation diagnostics and help prevent the selection of false positives. For example, the ACF is less sensitive to the harmonics of \prot\ than the wavelet analysis. The CS accentuates common peaks between the ACF and GWPS while attenuating signals that  correspond potentially to a harmonic of \prot\ or that are potentially not related to stellar rotation, for example instrumental artifacts.

 The rotation period estimates obtained from the full solar light curve are: $P_\text{GWPS}=22.87\pm3.47$ d; $P_\text{ACF}=26.57$ d, and $P_\text{CS}=26.43\pm1.04$ d. The $P_\text{GWPS}$ is short in comparison with the expected rotation at the sunspot latitudes, but the uncertainty is relatively large, and, thus, the expected value is within the error bars. The GWPS (panel b of Fig.~\ref{fig:rot_VIRGO}) shows a blended band of strong rotational signal ranging roughly from the first to the third harmonics of the rotation period. This results into three broad overlapping peaks in the GWPS. The amplitude of the peaks associated with $P_\text{ACF}$ and $P_\text{CS}$ is relatively small. In particular, the fast decay of the ACF suggests short-lived spots or active regions \citep{Lanza2014,Giles2017,Santos2021_tacf}, which is known to be the case of the majority of the solar active-regions. In Appendix~\ref{sec:appsun-like} we show that the results and comparison between the Sun and the Sun-like stars do not change significantly whether we adopt the $P_\text{GWPS}$ or $P_\text{CS}$. Nevertheless, particularly concerning the selection of Sun-like stars, we adopt $\text{P}_{\text{rot}\odot}=26.43$ d to compute the \sph.

\begin{figure}[h!]
    \centering
    \includegraphics[width=\hsize]{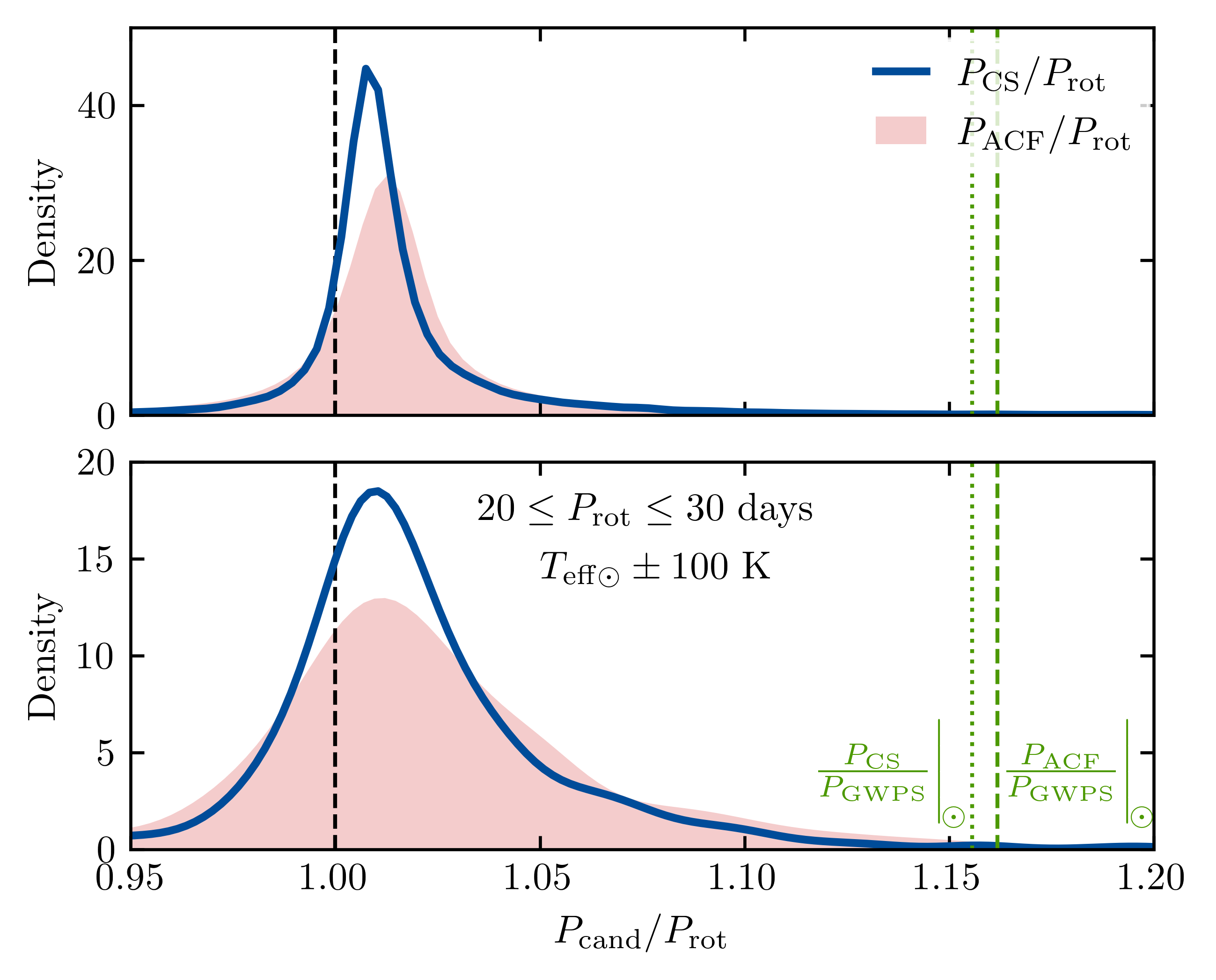}\vspace{-0.2cm}
    \caption{Distribution of the ratio between the $P_\text{ACF}$ (red shaded region) or $P_\text{CS}$ (solid blue line) and the final \prot\ for \kep\ solar-like stars \citep{Santos2019a,Santos2021ApJS}, where ``cand'' stands for period candidate. The top panel corresponds to all stars where $P_\text{cand}$ are within 30\% with respect to \prot. The bottom panel highlights the stars with \prot\ estimates within 20 and 30 days and \teff\ around $\text{T}_{\text{eff}\odot}$. The black dashed line marks $P_\text{cand}/P_\text{rot}=1$ and the green dashed and dotted lines indicate the $P_\text{ACF}/P_\text{GWPS}$ and $P_\text{CS}/P_\text{GWPS}$ for the Sun, respectively.}
    \label{fig:Prot_estimates}
\end{figure}

In relation to the rotational analysis of the \kep\ targets, there is a slight tendency for the GWPS to recover periods that are shorter than those retrieved through the ACF and CS (Fig.~\ref{fig:Prot_estimates}). Nevertheless, the difference tends to be small and is encompassed by the uncertainty associated with the GWPS period estimate. Figure~\ref{fig:Prot_estimates} shows the distribution of the ratio between the $P_\text{ACF}$ or $P_\text{CS}$ (where ``cand'' stands for candidate) and the final \prot\ from \citet{Santos2019a,Santos2021ApJS}, being that \prot\ mostly corresponds to the $P_\text{GWPS}$. Figure~\ref{fig:Prot_estimates} illustrates the reliability of the GWPS estimates in \citet{Santos2019a,Santos2021ApJS} as the difference between the different period estimates is on average small and differences as great as that of the Sun are not common. \vspace{-0.2cm}

This verification ensures that there is no bias related to \prot\ when selecting the Sun-like stars. Furthermore, we correct the \avsph\ and \stdsph\ for their dependences on different properties, including \prot. Still, in Appendix~\ref{sec:appsun-like}, we compare the Sun with a group of stars rotating with periods around $P_\text{GWPS}$. 

\begin{figure}[h]
    \centering
    \includegraphics[width=\hsize]{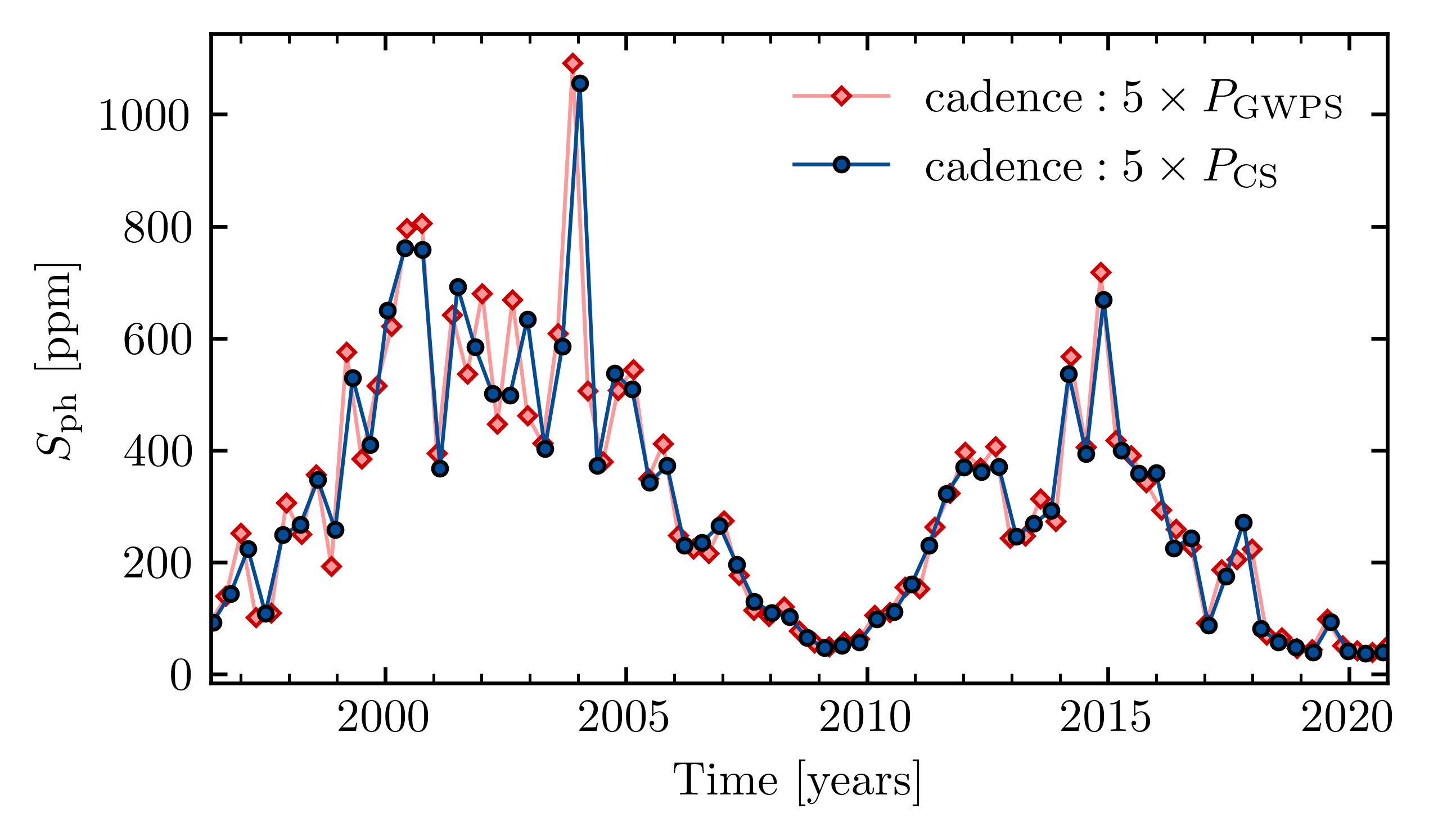}
    \caption{Comparison between the solar \sph\ computed when adopting the $P_\text{GWPS}$ (red) or $P_\text{CS}$ (blue). We use the latter as the reference $\text{S}_{\!\text{ph}\odot}$.}
    \label{fig:sphsun_comparison}
\end{figure}

We then compute the $\text{S}\!_{\text{ph}\odot}$ over segments of length $5\times \text{P}_{\text{rot}\odot}$. Figure~\ref{fig:sphsun_comparison} compares the \sph\ when adopting the $P_\text{GWPS}$ (red) or $P_\text{CS}$ (blue; adopted as $\text{P}_{\text{rot}\odot}$). Since the timescales are very similar, the differences are negligible.

For comparison and to have access to a more complete picture of the magnetic activity of the present Sun, here we also use the SAs and radio flux (\F) data, which we re-binned to the same cadence as $\text{S}\!_{\text{ph}\odot}$. The re-binning was done by computing the average SA or \F\ over a given time interval of length $5\times \text{P}_{\text{rot}\odot}$.

\begin{figure*}[t!]
    \centering
    \includegraphics[width=\hsize]{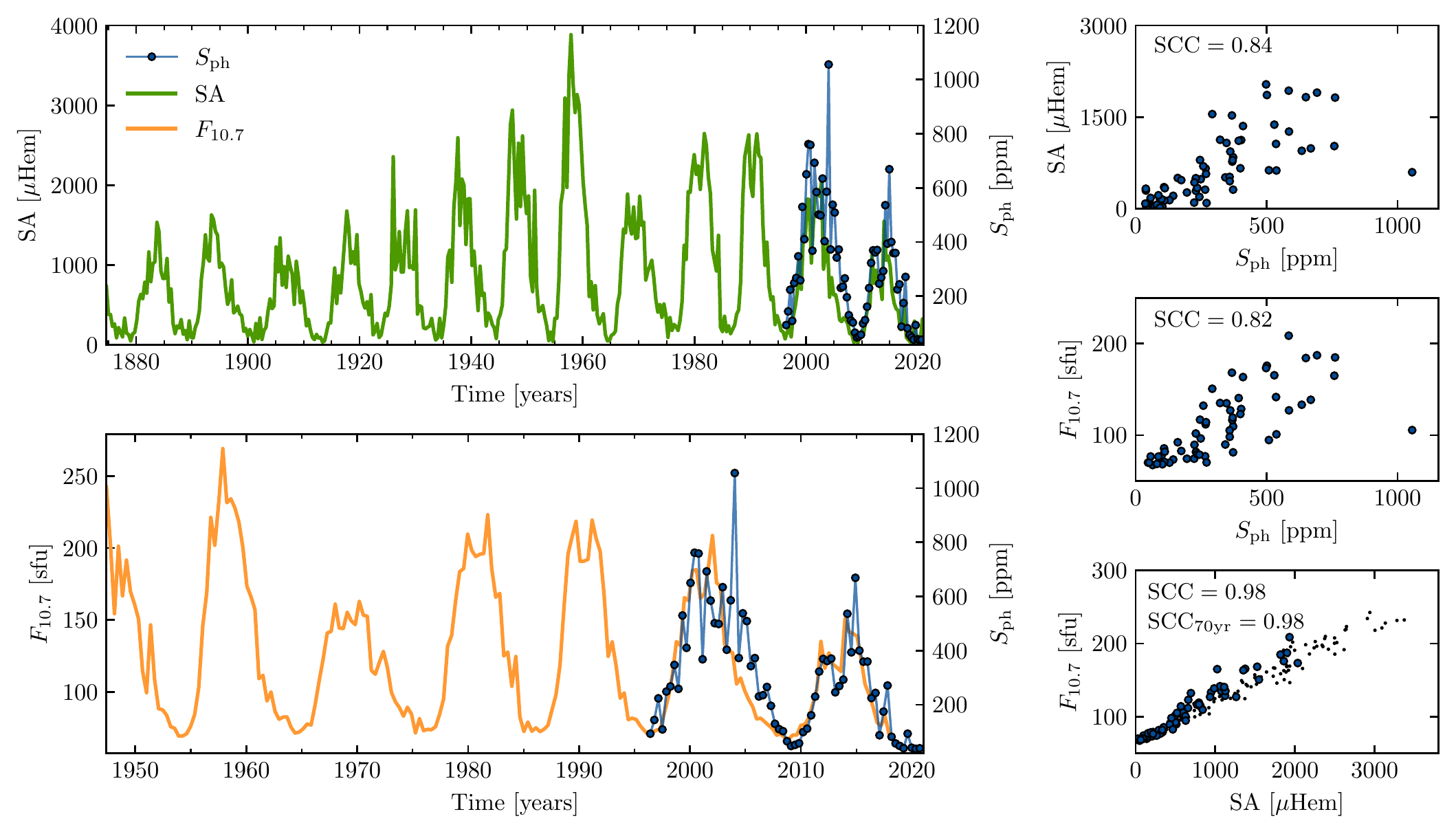}\vspace{-0.1cm}
    \caption{Comparison between the solar magnetic APs. In the left panels, \sph\ from VIRGO g+r is shown by the blue circles, while SA and \F\ are shown by the solid green and orange lines, respectively. The right-hand side shows how \F, SA, and \sph\ relate  to each other: the blue circles correspond to solar cycles 23 and 24, the black dots correspond to the $\sim\!70$ years of contemporaneous observations of SA and \F. Each panel shows the corresponding SCC value, where ``70yr'' stands for 70 years of common SA and \F\ observations.}
    \label{fig:solarproxies}
\end{figure*}

Figure~\ref{fig:solarproxies} compares the different magnetic APs with the cadence $5\times \text{P}_{\text{rot}\odot}$. It is clear that the re-binned SA (blue) and \F\ (orange) behave more smoothly than the $\text{S}\!_{\text{ph}\odot}$. The right-hand side panels of Fig.~\ref{fig:solarproxies} show how the different magnetic APs relate to each other for the contemporaneous observations. Each panel also lists the respective SCC. Due to the sensitivity of different APs to different layers of the atmosphere, and different phenomena related to magnetic activity that have different spatial distributions, magnetic hysteresis is observed as different proxies have different ``paths'' in the rising and declining phases of the cycle \citep[see, e.g.,][]{Jain2009,Salabert2017}. As illustrated in Fig.~\ref{fig:solarproxies}, the declining phase tends to be faster in the SA and \F\ than in the \sph. Nevertheless, the different APs are strongly correlated. Particularly, \citet{Salabert2017} found that $\text{S}\!_{\text{ph}\odot}$ best correlated to the SA.

\begin{figure}[h!]
    \centering
    \includegraphics[width=\hsize]{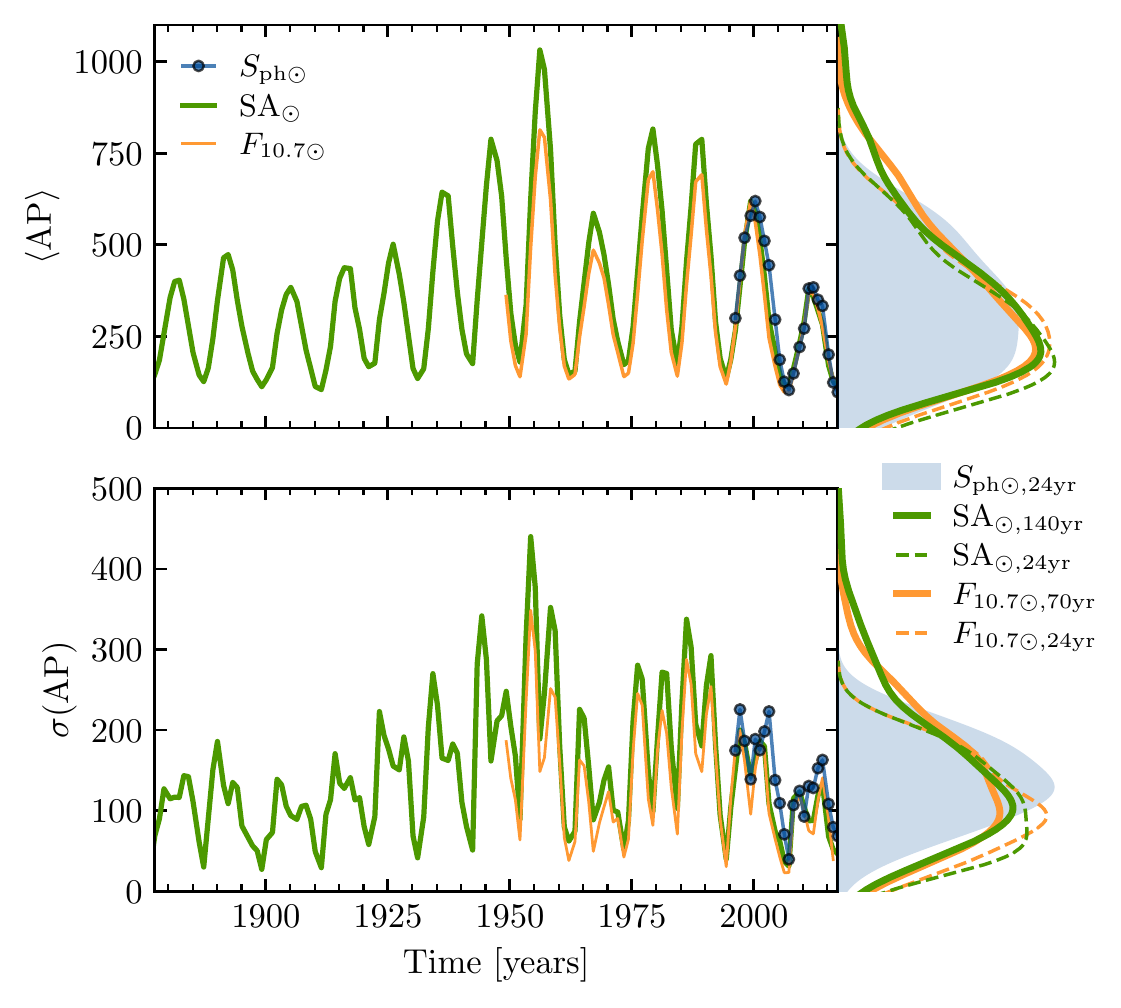}
    \caption{Comparison between the average (top) and standard deviation (bottom) values of the APs computed from 4-year segments. In the left panels, the blue symbols correspond to the $\text{S}\!_{\text{ph}\odot}$, while the green and orange solid lines concern the scaled SA and \F, respectively. The right-hand panels show their distributions. ``24yr'' denotes 24 years, i.e., the two last solar cycles, while ``70yr'' and ``140yr'' denote 70 and 140 years.}\vspace{0.1cm}
    \label{fig:scaledAPapp}
\end{figure}

We then split the solar data into 4 yr segments, spaced by one year: 22 $\text{S}\!_{\text{ph}\odot}$ segments (Sect.~\ref{sec:ressun}); 144 SA segments; and 68 \F\ segments. We computed the average SA and \F. We then scaled $\langle \text{SA}\rangle$ and $\langle F_{10.7}\rangle$ to the $\langle \text{S}\!_{\text{ph}\odot}\rangle$. The scaling is determined by the solar cycles 23 and 24\footnote{Solar cycle 23 has larger amplitude and deeper minimum than solar cycle 24. Thus, the scaling is in fact ruled by solar cycle 23.} and then applied to the remainder of the re-binned SA and \F\ time series. The respective standard deviation for each 4 yr segment is then computed from the scaled data. Both average and standard deviation values of the APs are shown in Fig.~\ref{fig:scaledAPapp}. For simplification, hereafter, we refer to the scaled SA and \F\ simply as SA and \F. As the SA and \F\ data were scaled using solar cycles 23 and 24, the amplitude of those cycles approximately match in the different APs. Interestingly, the amplitude of solar cycle 24 is similar in the scaled SA and \F. However, for the preceding cycles the amplitude in \F\ is systematically smaller than in SA. 

We note that the scaling of the APs could also be determined using the smoothed version of the AP by applying a boxcar filter, which is often considered, for example, to remove or isolate short-term variations, namely the QBOs. Similarly to what is seen for the $\text{S}\!_{\text{ph}\odot}$ in Sect.~\ref{sec:ressun}, the signature of the QBO is present in the 4 yr $\sigma$(AP) around the activity maxima (bottom panel of Fig.~\ref{fig:scaledAPapp}).

The right-hand panels of Fig.~\ref{fig:scaledAPapp} show the AP distributions. The dashed lines correspond to the SA and \F\ data solely for solar cycles 23 and 24. While several of the preceding cycles have amplitudes comparable with solar cycles 23 and 24, the top panels also illustrate the fact that the long SA and \F\ records give access to larger amplitude cycles, that is to say, higher activity levels than those seen in the past 24 years. In addition to the signature of the QBO, the variation in the 4 yr AP data also accounts for the 11-year cycle. The distributions of the $\langle \text{SA}\rangle$ and 
$\langle F_{10.7} \rangle$ are shifted toward smaller values (a decrease of $\sim13\%$ in comparison with the $\langle \text{S}\!_{\text{ph}\odot}\rangle$). The slight difference is due to magnetic hysteresis, particularly the declining phases in SA are faster than in $\text{S}\!_{\text{ph}\odot}$ \citep[see also][]{Salabert2017}. However, the bias introduced when re-binning SA and \F\ is clear in the comparison between the $\sigma$(AP) (bottom). 

%\clearpage
%\pagebreak

\section{\textit{Kepler} target sample}\label{sec:cleansample}

\subsection{Spectral types and evolutionary stage}

\begin{figure}[h!]
    \centering
    \includegraphics[width=\hsize]{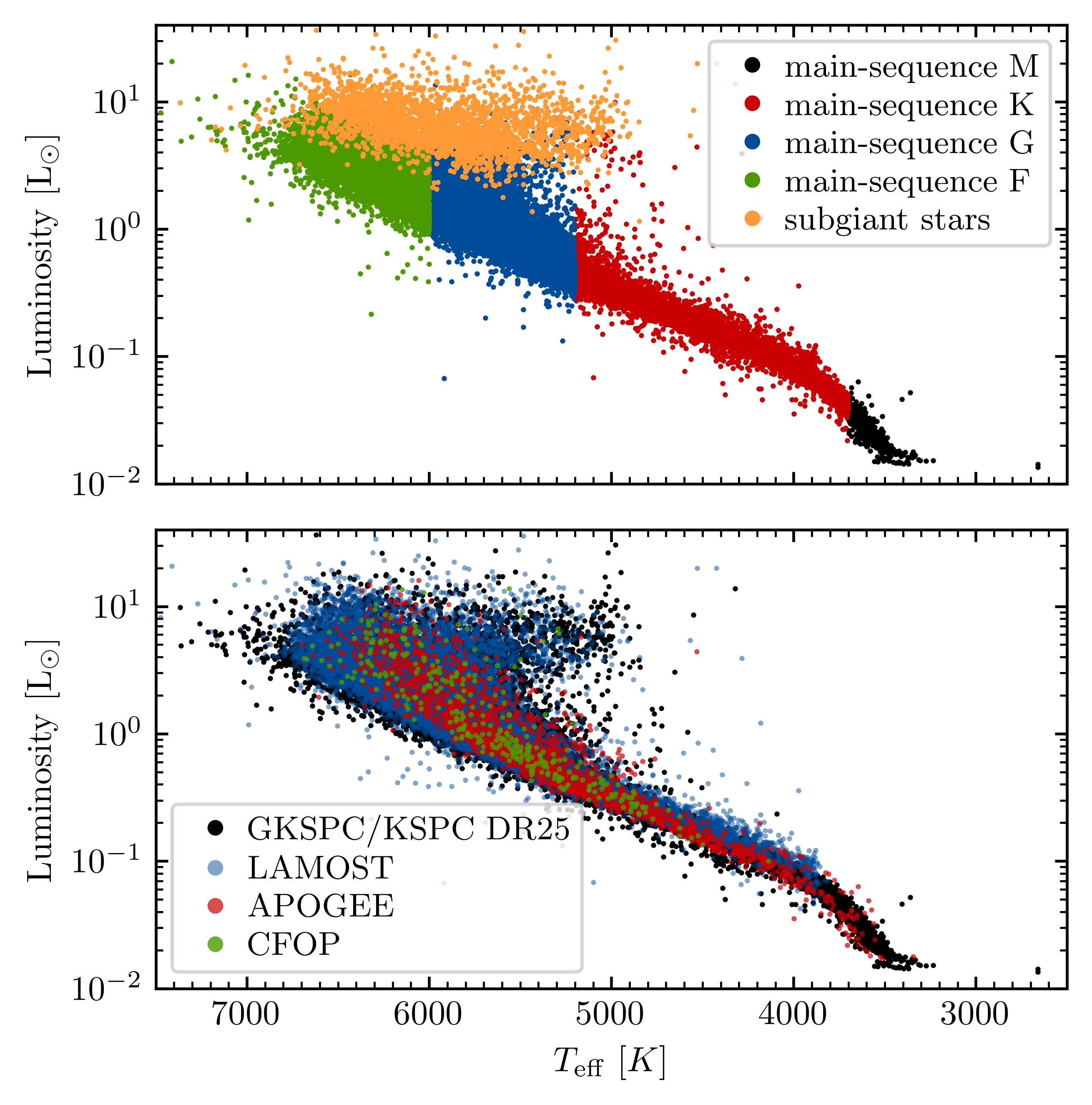}
    \caption{HR diagram for the target sample. The color code indicates to which subsample the targets belong to (top) and the source for the stellar properties (bottom).}\vspace{0.5cm}
    \label{fig:hr}
\end{figure}

We split our main sample into subsamples, according to the spectral type and possible evolutionary stage. We adopted the same \teff\ cuts for the spectral types as in \citet{Santos2019a,Santos2021ApJS}: at 3700 K, 5200 K, and 6000 K. To split main-sequence and subgiant stars, we adopted the EEP values from {\color{blue2}Mathur et al. (in prep)}. The subsamples are illustrated in Fig.~\ref{fig:hr} together with the stellar properties source. 

\subsection{Removing potential sources of bias}\label{app:clean}

As described in Sect.~\ref{sec:sample}, we have neglected targets that may contribute to a bias on our analysis. Here, we show where those targets lie on the \stdsph\ versus \avsph\ diagram.

\begin{figure}[h]
    \centering
    \includegraphics[width=\hsize]{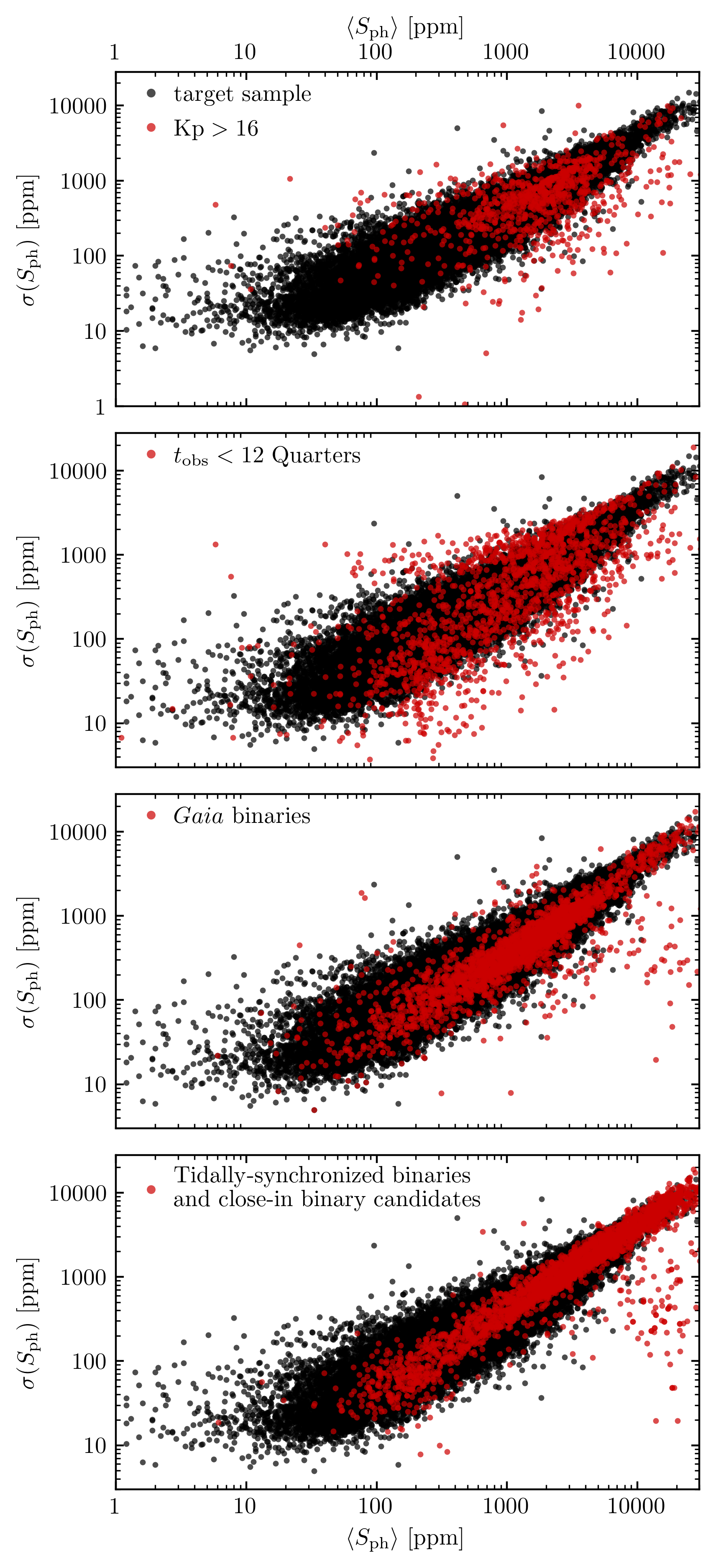}
    \caption{\stdsph\ as a function of the \avsph. The black symbols show the target sample of the current study, while the red symbols show the targets that were removed according to a given criteria: \kep\ magnitude, observational length, and potential binarity.}
    \label{fig:cleansph}
\end{figure}

We neglected targets with a light curve shorter than 12 \kep\ Quarters. The shorter is the light curve, the less complete is the information retrieved on the photometric activity. This selection criterion removes some of the outliers (above and below) in the \stdsph\ versus \avsph\ relation (top panel of Fig.~\ref{fig:cleansph}). 

The light curves of faint stars tend to be noisy. For faint stars, the spot rotational modulation becomes harder to detect and its amplitude, related to the \sph, is affected. For that reason, we removed for this analysis targets with $\text{Kp}>16$. This selection criterion also removed a significant number of outliers (second panel of Fig.~\ref{fig:cleansph}), particularly those lying below the general trend.

Magnetic activity can be affected by binarity. Thus, in addition to all known eclipsing binaries, which were already absent from the rotational analysis \citep{Santos2019a,Santos2021ApJS}, we removed all binary candidates identified by \citet[][{\it Gaia} binaries]{Berger2018,Binaries_GaiaDR3}, by \citet[][tidally synchronized binaries]{Simonian2019}, and by \citet[close-in binary candidates]{Santos2019a,Santos2021ApJS}. This removed a significant number of targets that follow the \stdsph\ versus \avsph\ relationship (third and bottom panels of Fig.~\ref{fig:cleansph}), but it also removed many outliers in particular with smaller \stdsph\ or larger \avsph\ than expected.

While removing a large number of targets that follow the \stdsph\ versus \avsph\ relation, it becomes clear that the majority of the outliers were eliminated by removing these potential contaminants. Thus, all the targets in red in Fig.~\ref{fig:cleansph} were neglected from the analysis. 

\begin{figure}[h]
    \centering
    \includegraphics[width=\hsize]{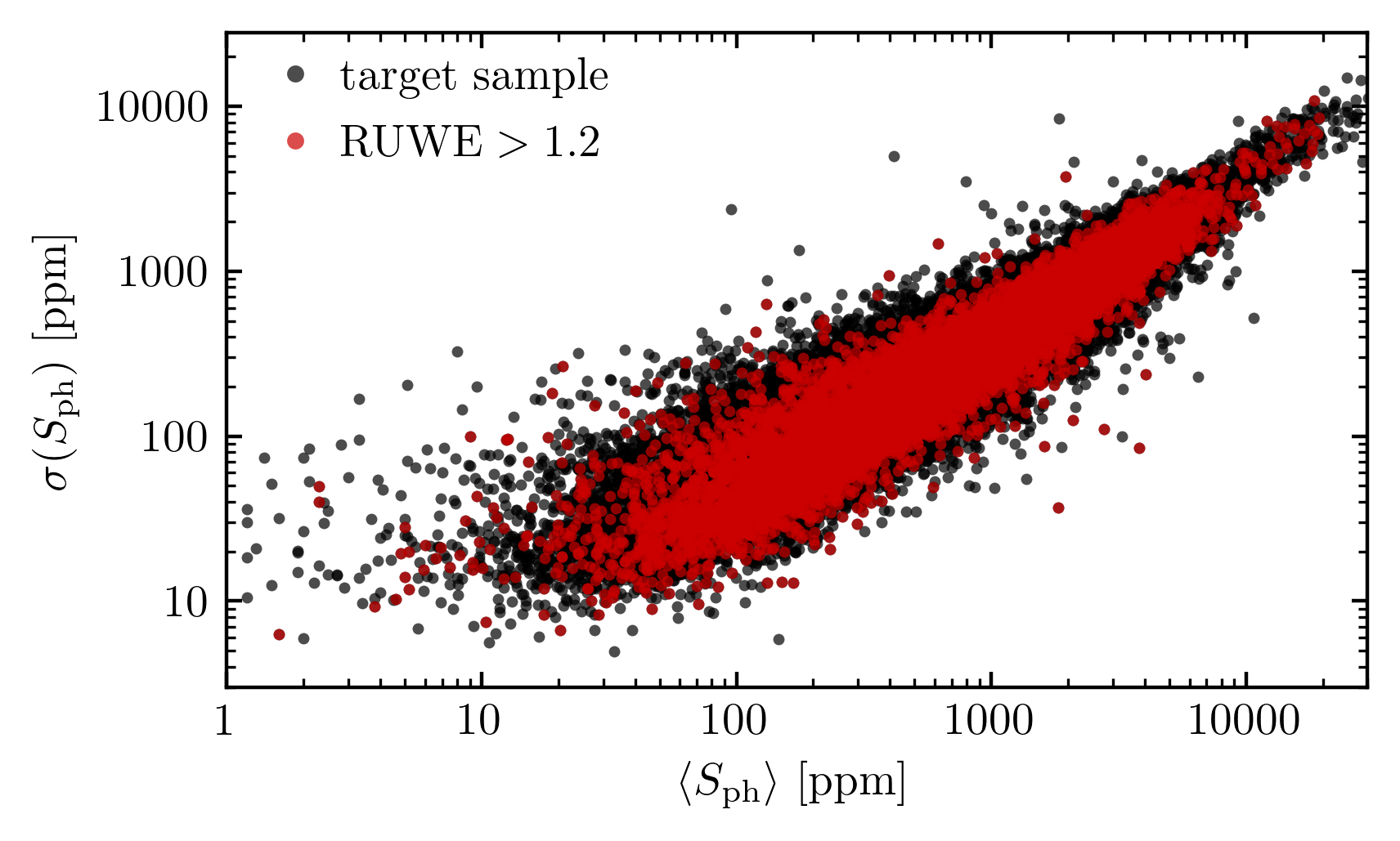}
    \caption{\stdsph\ as a function of the \avsph\ for the target sample, where red highlights the targets with $\text{RUWE}>1.2$.}
    \label{fig:ruwe}
\end{figure}

Finally, we also checked whether considering an additional selection criterion based on the RUWE from {\it Gaia} Early Data Release 3 \citep[EDR3;][]{GaiaDR1,GaiaEDR3,Lindegren2021} changes the results. Targets with $\text{RUWE}>1.2$ are often considered to be likely binaries. Our target sample has 6,464 targets with $\text{RUWE}>1.2$, but we do not find that the targets with large RUWE are outliers in terms of \sph\ and \prot\ behavior. For this reason we decided to keep them in the target sample. In Table~\ref{tab:ruweleq}, we summarize the results when neglecting these targets and show that the results and conclusions do not change significantly. Figure~\ref{fig:ruwe} shows where those targets lie in the \stdsph\ versus \avsph\ diagram. 

\begin{figure}[h]
    \centering
    \includegraphics[width=\hsize]{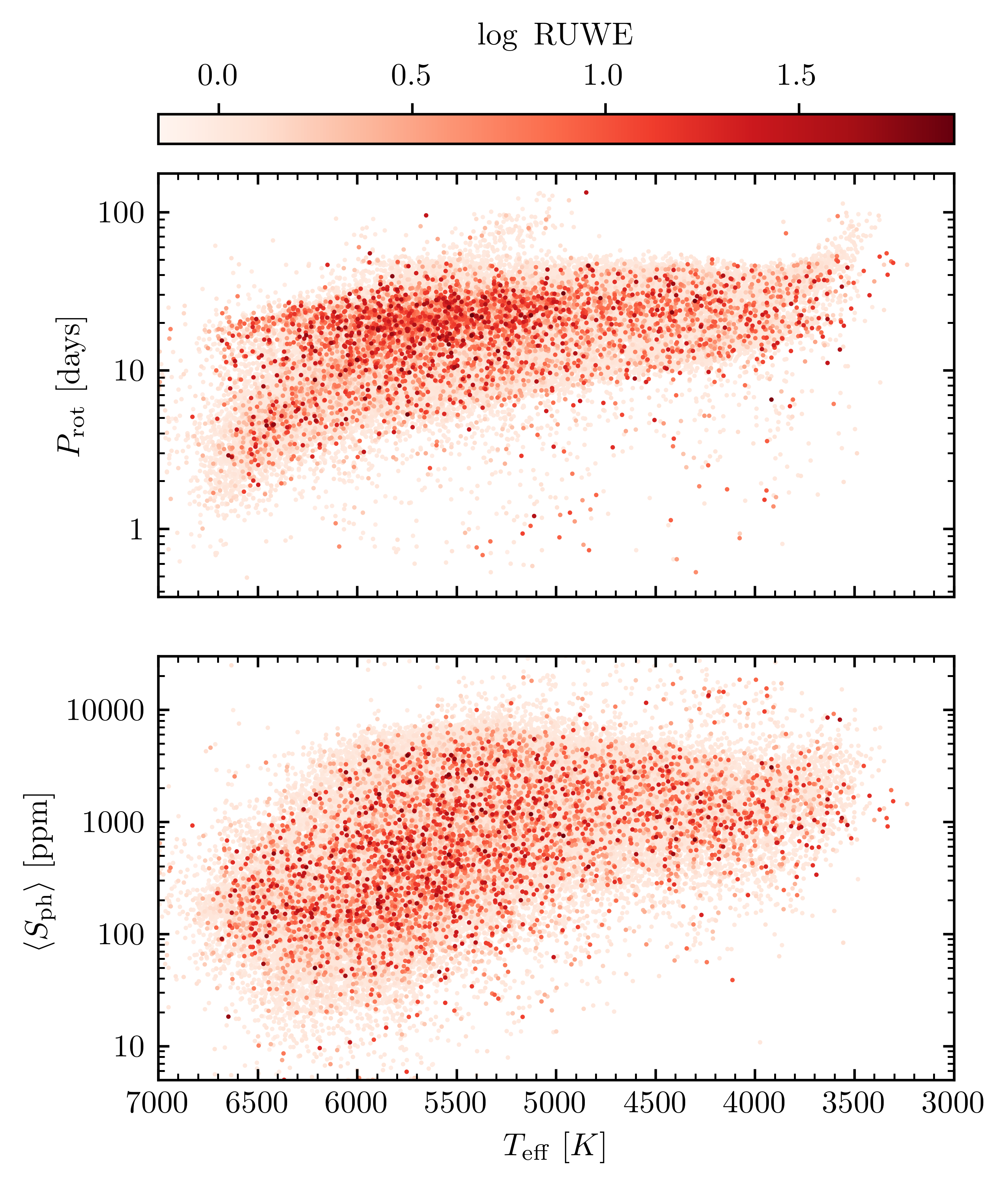}
    \caption{\prot\ (top) and \avsph\ (bottom) as a function of the effective temperature for the stars in the target sample, including the subgiant stars. The color code indicates the RUWE values for each target.}
    \label{fig:ruwe2}
\end{figure}

Figure~\ref{fig:ruwe2} shows the rotation versus \teff\ and activity versus \teff\ diagrams color-coded by RUWE. The data points are ordered from the smallest (light red) to the largest (dark red) RUWE values, such that the latter overlap the former for better visualization of the targets with large RUWE. Most of the targets with large RUWE are within the normal parameters. There is a large concentration of dark data points in the location of the slow-rotating population, which could be expected because it is the most numerous population.

%\pagebreak

\subsection{Bimodal rotation-period distribution: Slow- and fast-rotating populations}\label{sec:approt}

The distribution of rotation periods for \kep\ targets is bimodal \citep[e.g.,][]{McQuillan2014,Santos2019a,Santos2021ApJS}, particularly for the cool solar-like stars. The bimodality of \prot\ leads to two populations or groups in the \prot\ versus \teff\ diagram with a region of low density in between (also called intermediate-\prot\ gap). The discontinuity in the rotation distribution can also be seen in the activity-rotation diagrams, where two regimes seem to exist. While less pronounced for G dwarfs, \sph\ depends on \prot\ in both fast- and slow-rotating populations. 

We used the discontinuity in the activity-rotation relation to split the GKM dwarfs into fast- and slow-rotating populations. While our samples, metrics, and approach differ from those in \citet{Reinhold2020}, they also used these discontinuities to determine the location of the ``gap'' for solar-like stars observed during K2 \citep{Howell2014}.

\begin{figure}
    \centering
    \includegraphics[width=\hsize]{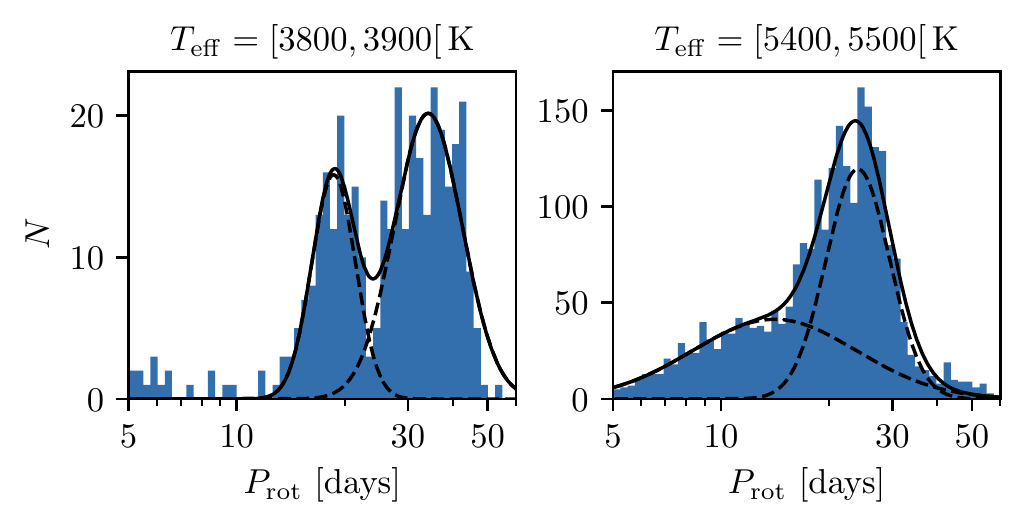}\vspace{-0.2cm}
    \caption{Rotation-period distribution for stars with \teff\ within 3800 and 3900 K (left) and 5400 and 5500 K (right). The solid line shows the best fit with a double Gaussian, while the dashed lines show the individual Gaussian distributions.}
    \label{fig:bimodal}\vspace{-0.2cm}
\end{figure}

First, we take \teff\ intervals of 100 K from 3500 K to 6000 K (only a few stars are cooler than 3500 K). Figure~\ref{fig:bimodal} shows two examples of the rotation-period distribution in such narrow intervals. While the bimodality of the \prot\ distribution is clearer for the K and M dwarfs, the log~\prot\ distribution for G dwarfs is also well described by a double Gaussian. 

\begin{figure}
    \centering
    \includegraphics[width=\hsize]{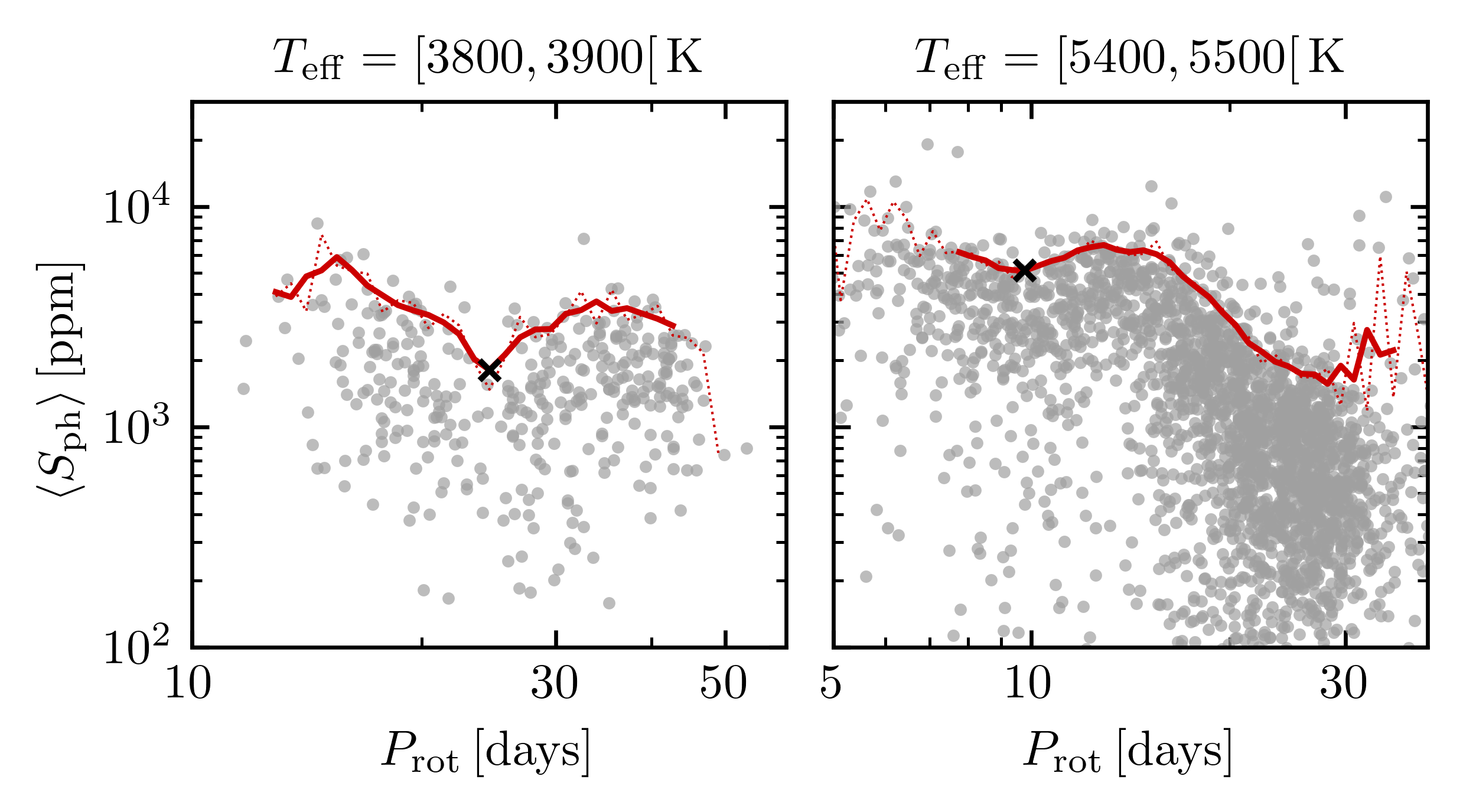}\vspace{-0.2cm}
    \caption{Activity-rotation diagram for two intervals of 100 K: 3800-3900 K (left); and 5400-5500 K (right). We use the local minimum (black crosses) to split the sample into fast- and slow-rotating populations. The gray dots show the stars with \teff\ within the considered interval. The dotted red line shows the upper edge of the \avsph\ distribution (95\textsuperscript{th} percentile), while the solid red line shows its smoothed version.}
    \label{fig:split-method}
\end{figure}

For each \teff\ interval, we determined the upper edge of the activity-rotation diagram by: 1) splitting the data now into log~\prot\ intervals of width 0.02 dex; and 2) computing the 95\textsuperscript{th} percentile for the \avsph\ in each interval (dotted red line in Fig.~\ref{fig:split-method}). We then smoothed the upper edge with an uniform filter of size 3 (solid red) and determine the location of the local minimum between the two populations (black cross; through bounded minimization). For the minimization procedure, we neglected the regions with few stars, where the upper edge is not well defined. Figure~\ref{fig:split-method} shows two examples, for the same \teff\ intervals as in Fig.~\ref{fig:bimodal}.

\begin{figure}
    \centering
    \includegraphics[width=\hsize]{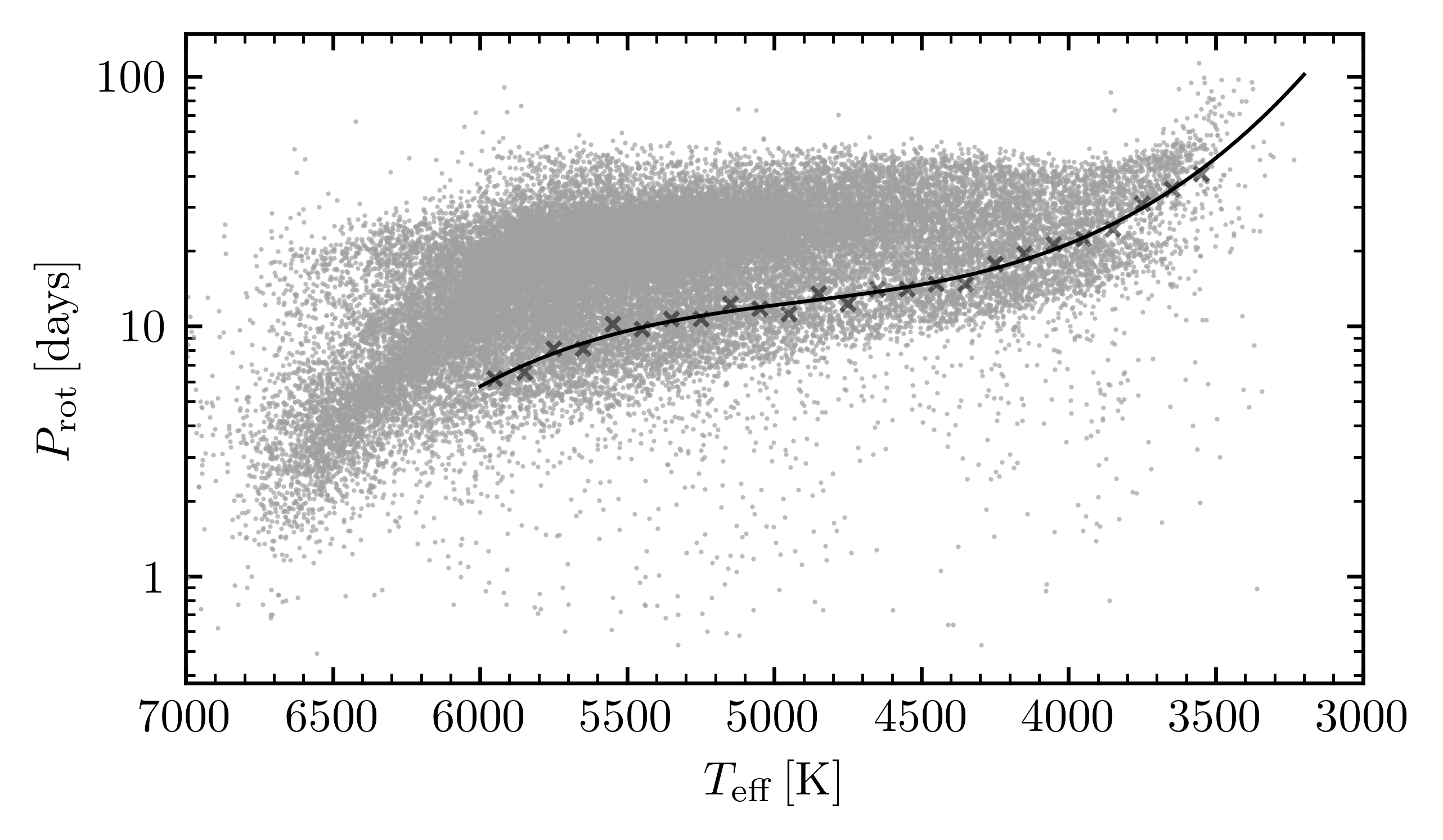}
    \caption{Rotation period as a function of the effective temperature for the main-sequence stars in our target sample. The black crosses show the position of the local minimum in the \avsph-\prot\ diagram for each \teff\ interval of 100 K (Fig.~\ref{fig:split-method}). The black solid line shows the best fit with a third degree polynomial, which is adopted in this work to split the slow- and fast-rotating populations.}
    \label{fig:branches_teff}
\end{figure}

Figure~\ref{fig:branches_teff} shows the rotation period as a function of \teff\ for the main-sequence stars in our sample. The crosses mark the local minima found in the previous step (Fig.~\ref{fig:split-method}) and the solid line corresponds to the best fit to the local minima with a third degree polynomial. We adopted this polynomial as the transition between the two regimes for GKM dwarfs. 

\subsection{Sun- and Doris-like stars}\label{sec:appsun-like}

In Sect.~\ref{sec:sunlike}, we select stars with very similar properties (\teff, $L$, and \prot) to those of the Sun (i.e., Sun-like stars). Since the work by \citet{Reinhold2020Sci}, the stellar properties have been updated \citep{Ahumada2020,Zong2020,Berger2020,Santos2021ApJS}. In this subsection we provide more details on the Sun-like stars in our sample and their previous classification. We also compare the properties of the Sun- and Doris-like stars in comparison with the G dwarf sample.

\begin{figure}[h]
    \centering
    \includegraphics[width=\hsize]{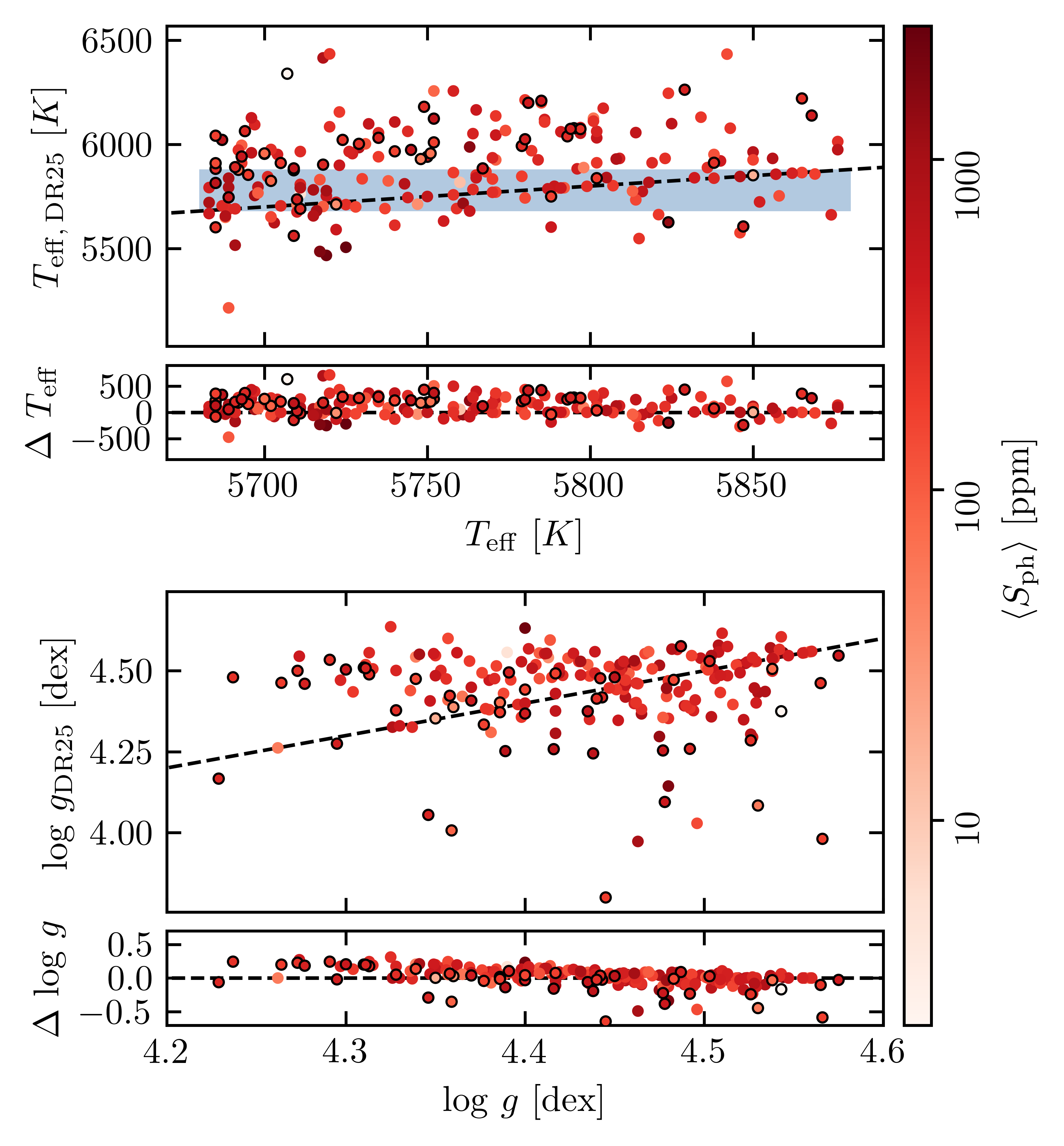}
    \caption{Comparison between the current values for \teff\ and \logg\ (x-axis) and those from KSPC DR25 (y-axis) for the Sun-like sample. $\Delta\,T_\text{eff}$ and $\Delta\,\log\,g$ correspond to the difference between previous and current constraints. The data points are colored according to their \avsph. The dashed lines mark the 1-1 lines and the zero difference. The blue shaded region marks the \teff\ interval we consider in this study to select Sun-like stars.}
    \label{fig:teff_logg_sun}
\end{figure}

The Sun-like sample is composed of 211 stars (Sect.~\ref{sec:sunlike}). These are stars within 100 K around $\text{T}_{\text{eff}\odot}$, 0.3 around $\text{L}_\odot$, and 2 days around $\text{P}_{\text{rot}\odot}$. In terms of rotation-period estimates, from the 211 stars, 43 are in the periodic sample of \cite{McQuillan2014} and 158 in their nonperiodic sample. 10 stars were not part of their analysis. Figure~\ref{fig:teff_logg_sun} compares the current \teff\ and \logg\footnote{We opted to use luminosity instead of \logg\ in our analysis. Nevertheless, the final results are consistent whether we adopt $L$ or \logg.} values with those available previously from KSPC DR25 (simply DR25 in Fig.~\ref{fig:teff_logg_sun}). Many of the targets classified in this work as Sun-like had KSPC DR25 properties inconsistent with Sun-like. In KSPC DR25, most targets were considered to be hotter than current constraints. The average difference between the previous and current \teff\ values is 129.1 K, while the for \logg\ is 0.01 dex.

\begin{figure}[h]
    \centering
    \includegraphics[width=\hsize]{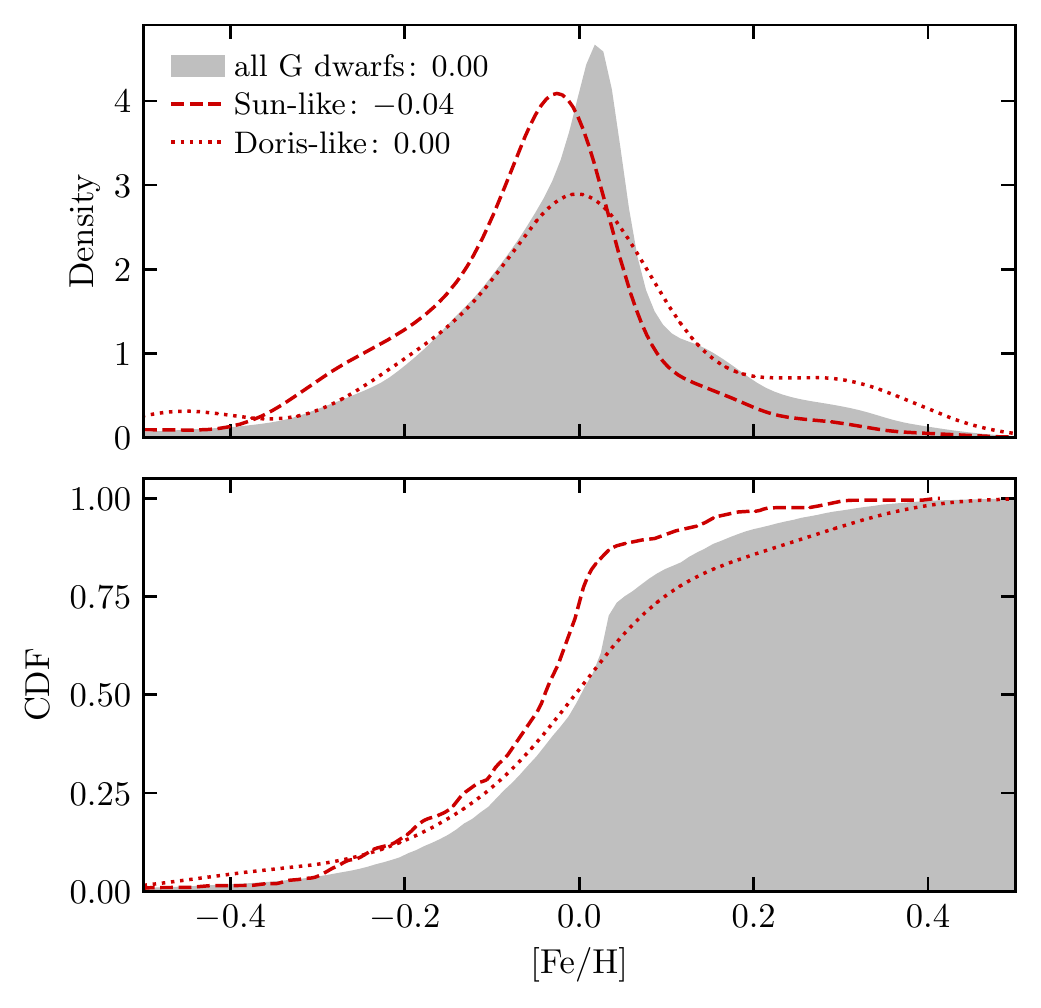}
    \caption{\metal\ distribution for the full G-dwarf sample (shaded gray), the Sun-like stars (dashed red), and the Doris-like stars (dotted red). The bottom panel shows the CDFs, where one can identify more clearly the differences between the samples. The median values of the distributions are also identified.}
    \label{fig:metal_sun_doris_app}
\end{figure}

As noted in Sects.~\ref{sec:dorislike} and \ref{sec:metal} the Doris-like stars are more metallic than what would be expected for the \kep\ field. Figure~\ref{fig:metal_sun_doris_app} shows the \metal\ distributions for the full G-dwarf sample (shaded gray), the Sun-like stars (dashed red), and the Doris-like stars (dotted red). The median values are indicated in the top figures. For the Sun-like sample, there is an excess of stars with solar metallicity in comparison with the distribution for the full G-dwarf sample. For the Doris-like sample, within that \teff\ and $L$ range and with such \prot, there is a large number of high metallicity stars.

\begin{figure}[h]
    \centering
    \includegraphics[width=\hsize]{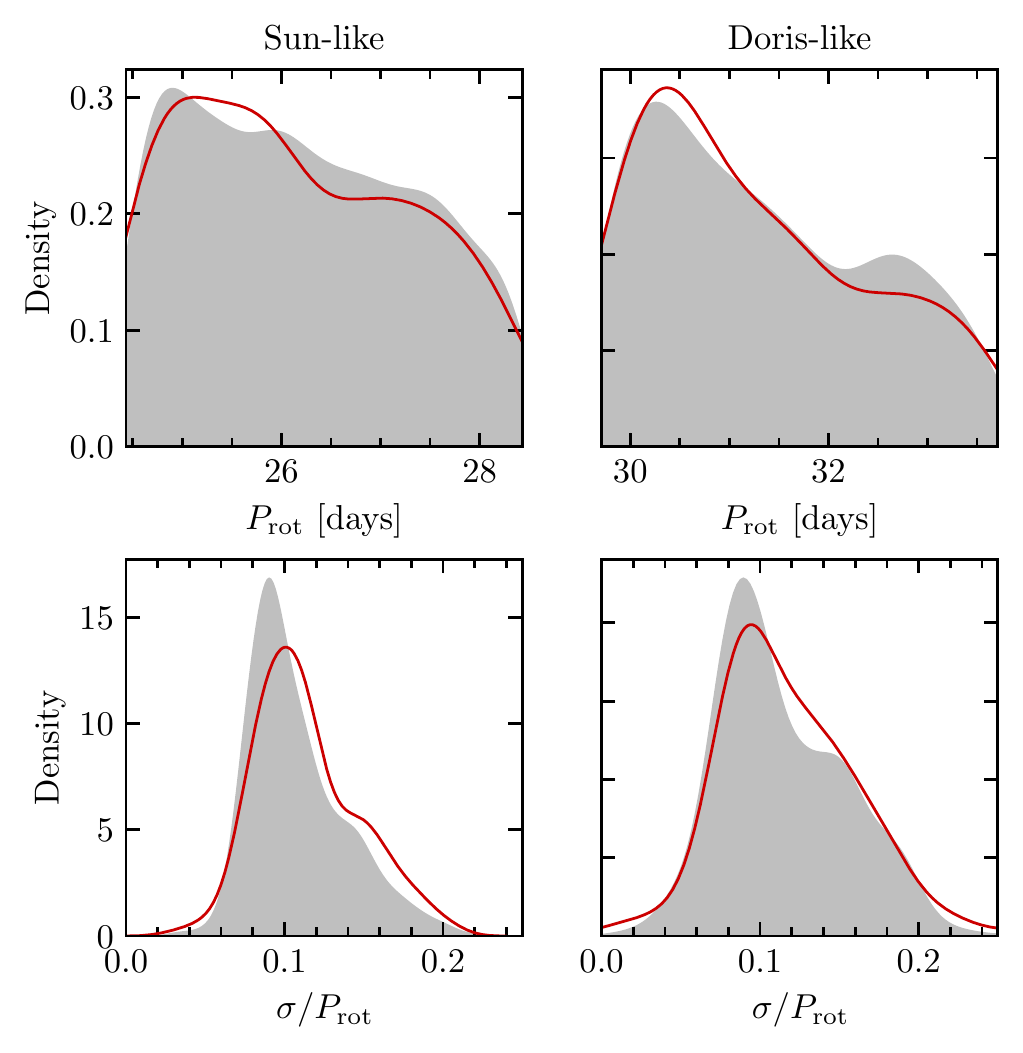}
    \caption{Comparison between the distributions of \prot\ (top) and relative uncertainty (bottom) for the G dwarfs with \prot\ within $\text{P}_{\text{rot}\odot}\pm 2$ (shaded gray) and the Sun-like (left) and Doris-like (right) samples (red).}
    \label{fig:prot_sun_doris}
\end{figure}

Figure~\ref{fig:prot_sun_doris} shows the distributions of \prot\ and the relative uncertainty for Sun- (left) and Doris-like (right) samples in comparison with the distribution for the G~dwarfs with similar \prot\ (shaded gray; full \teff\ range, i.e., within 5200 and 6000 K). The \prot\ distributions show a similar behavior, with a higher fraction of detections for shorter \prot\ than for longer \prot\ in this parameter space. Therefore, the bias toward shorter \prot\ is expected in the Sun- and Doris-like samples. The \prot\ uncertainty is in average $\sim10\%$, which is consistent with the values for the full \kep\ sample \citep{Santos2021ApJS}.

\begin{figure}
    \centering
    \includegraphics[width=\hsize]{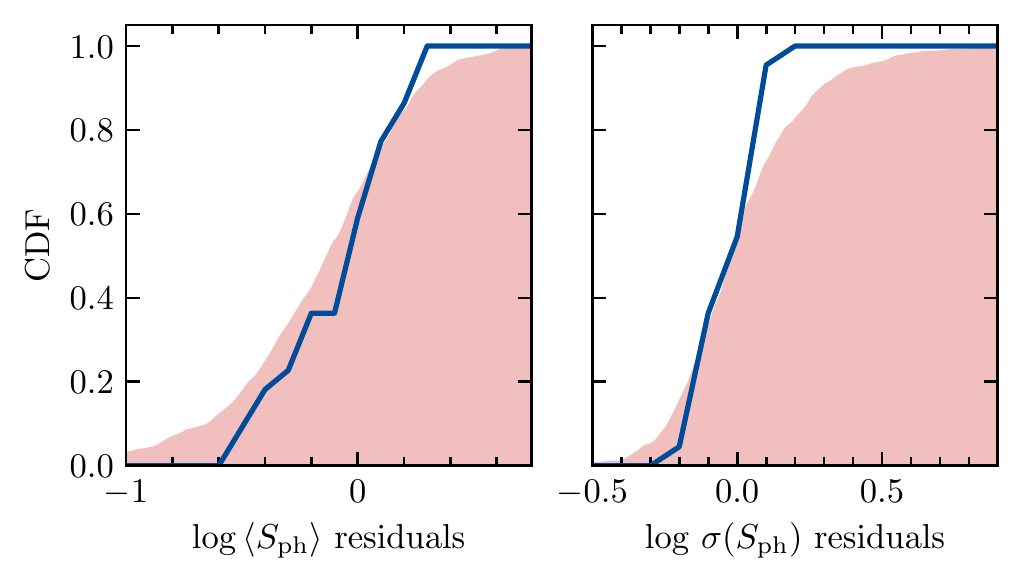}\vspace{-0.3cm}
    \caption{Same as in Fig.~\ref{fig:dist_avsph} but when considering the period estimate from the GWPS.}\vspace{-0.1cm}
    \label{fig:sunlike_P23}
\end{figure}

As seen in Appendix~\ref{sec:appsun}, there is some discrepancy between the period estimates from the rotational analysis of VIRGO g+r. The estimates agree within the error bars, but the $P_\text{GWPS}=22.87$ d underestimates $\text{P}_{\text{rot}\odot}$. Similarly to Fig.~\ref{fig:dist_avsph}, Fig.~\ref{fig:sunlike_P23} compares the CDF of the \avsph\ and \stdsph\ residuals for the Sun and the stars with similar \teff\ and $L$ to the Sun and \prot\ within $P_\text{GWPS}\pm2$ d. This selection adds more low-activity stars in comparison to the Sun-like stars but the distributions are comparable.

%\pagebreak
\section{Multivariate linear regression}\label{app:regression}

\begin{table*}[htp]\small
    \centering
    \begin{tabular}{r|cccc|c}
    \hline\hline
    & \multicolumn{4}{c|}{main sequence} & \\
    & M & K & G & F & subgiant \\\hline

    $c_0$ & -0.61 & 0.16 & 0.68 & 1.24 & 0.97\\
    $c_\text{Kp}$ & $1.0\times10^{-2}$ & $1.3\times10^{-2}$ & $2.7\times10^{-2}$ & $5.2\times10^{-2}$ & $4.4\times10^{-2}$ \\
    $c_{t_\text{obs}}$ & $-5.6\times10^{-3}$ & $-1.5\times10^{-2}$ & $-2.4\times10^{-2}$ & $-5.2\times10^{-2}$ & $-4.3\times10^{-2}$ \\
    $c_{T_\text{eff}}$ & $1.8\times10^{-4}$ & $3.4\times10^{-5}$ & $-4.1\times10^{-5}$ & $-1.1\times10^{-4}$ & $-1.1\times10^{-4}$ \\
    $c_{L}$ & $1.7\times10^{0}$ & $-6.5\times10^{-3}$ & $-2.0\times10^{-2}$ & $-1.4\times10^{-2}$ & $8.8\times10^{-4}$ \\
    $c_\text{[Fe/H]}$ & $4.3\times10^{-2}$ & $3.0\times10^{-2}$ & $1.6\times10^{-2}$ & $-8.1\times10^{-2}$ & $-3.9\times10^{-2}$ \\
    $c_{P_\text{rot}}$ & $-4.9\times10^{-3}$ & $-6.2\times10^{-3}$ & $-8.4\times10^{-3}$ & $-4.6\times10^{-3}$ & $-2.4\times10^{-3}$ \\
    $c_{\log\,\langle S_\text{ph}\rangle}$ & $8.8\times10^{-1}$ & $8.2\times10^{-1}$ & $7.8\times10^{-1}$ & $7.2\times10^{-1}$ & $7.8\times10^{-1}$ \\\hline

    \multicolumn{6}{c}{} \\
    \multicolumn{6}{l}{\textsc{Fast-rotating population}}\\\hline
    $c_0$ & 0.12 & 0.38 & 0.84 & \multicolumn{2}{c}{}\\
    $c_\text{Kp}$ & $3.9\times10^{-3}$ & $1.1\times10^{-2}$ & $7.0\times10^{-3}$ & \multicolumn{2}{c}{} \\
    $c_{t_\text{obs}}$ & $-1.4\times10^{-2}$ & $-1.8\times10^{-2}$ & $-2.5\times10^{-2}$ & \multicolumn{2}{c}{} \\
    $c_{T_\text{eff}}$ & $1.4\times10^{-4}$ & $-1.2\times10^{-5}$ & $-3.0\times10^{-5}$ & \multicolumn{2}{c}{} \\
    $c_{L}$ & $-1.7\times10^{0}$ & $-2.2\times10^{-3}$ & $-7.1\times10^{-3}$ & \multicolumn{2}{c}{} \\
    $c_\text{[Fe/H]}$ & $4.2\times10^{-2}$ & $3.2\times10^{-2}$ & $3.0\times10^{-2}$ & \multicolumn{2}{c}{} \\
    $c_{P_\text{rot}}$ & $-6.5\times10^{-3}$ & $-1.1\times10^{-2}$ & $-1.4\times10^{-2}$ & \multicolumn{2}{c}{} \\
    $c_{\log\,\langle S_\text{ph}\rangle}$ & $8.2\times10^{-1}$ & $8.5\times10^{-1}$ & $8.1\times10^{-1}$ & \multicolumn{2}{c}{} \\\hline

    \multicolumn{6}{c}{} \\
    \multicolumn{6}{l}{\textsc{Slow-rotating population}}\\\hline
    $c_0$ & -2.86 & 0.23 & 0.65 & \multicolumn{2}{c}{} \\
    $c_\text{Kp}$ & $2.1\times10^{-2}$ & $1.3\times10^{-2}$ & $2.9\times10^{-2}$ & \multicolumn{2}{c}{} \\
    $c_{t_\text{obs}}$ & $1.4\times10^{-3}$ & $-1.4\times10^{-2}$ & $-2.4\times10^{-2}$ & \multicolumn{2}{c}{} \\
    $c_{T_\text{eff}}$ & $6.7\times10^{-4}$ & $2.7\times10^{-5}$ & $-3.7\times10^{-5}$ & \multicolumn{2}{c}{} \\
    $c_{L}$ & $5.2\times10^{0}$ & $-6.2\times10^{-3}$ & $-2.7\times10^{-2}$ & \multicolumn{2}{c}{} \\
    $c_\text{[Fe/H]}$ & $1.7\times10^{-1}$ & $3.2\times10^{-2}$ & $2.1\times10^{-2}$ & \multicolumn{2}{c}{} \\
    $c_{P_\text{rot}}$ & $-4.4\times10^{-3}$ & $-6.9\times10^{-3}$ & $-8.4\times10^{-3}$ & \multicolumn{2}{c}{} \\
    $c_{\log\,\langle S_\text{ph}\rangle}$ & $9.6\times10^{-1}$ & $8.1\times10^{-1}$ & $7.7\times10^{-1}$ & \multicolumn{2}{c}{} \\\hline\hline

    \end{tabular}
    \caption{Coefficients from the multivariate regression summarized in Table~\ref{tab:residuals}, where $c_0$ is the constant term.}
    \label{tab:residuals-coefficients}\vspace{-0.6cm}
\end{table*}

To account for possible dependences on different stellar properties and isolate each dependence, we perform multivariate linear regressions. In the main text we summarize the main results and in this section we provide the regression coefficients and parameters for other subsets of the main sample.

Tables~\ref{tab:residuals-coefficients}- \ref{tab:spec} show the regression coefficients related to Tables~\ref{tab:residuals}, \ref{tab:sunlike}, and \ref{tab:dorislike}.
Table~\ref{tab:ruweleq} lists the correlation coefficients and the multivariate regression coefficients when neglecting targets with RUWE>1.2 (Sect.~\ref{app:clean}). The results and conclusions do not change significantly. The same is found when considering solely the targets with spectroscopic atmospheric parameters (Table~\ref{tab:spec}). For M dwarfs, the results change significantly, with moderate correlations found between the \sph\ variation and \teff, $L$, and \metal. However, the number of available targets is very small and may be the cause for such results. 

\begin{figure}[h!]
    \centering
    \includegraphics[width=\hsize]{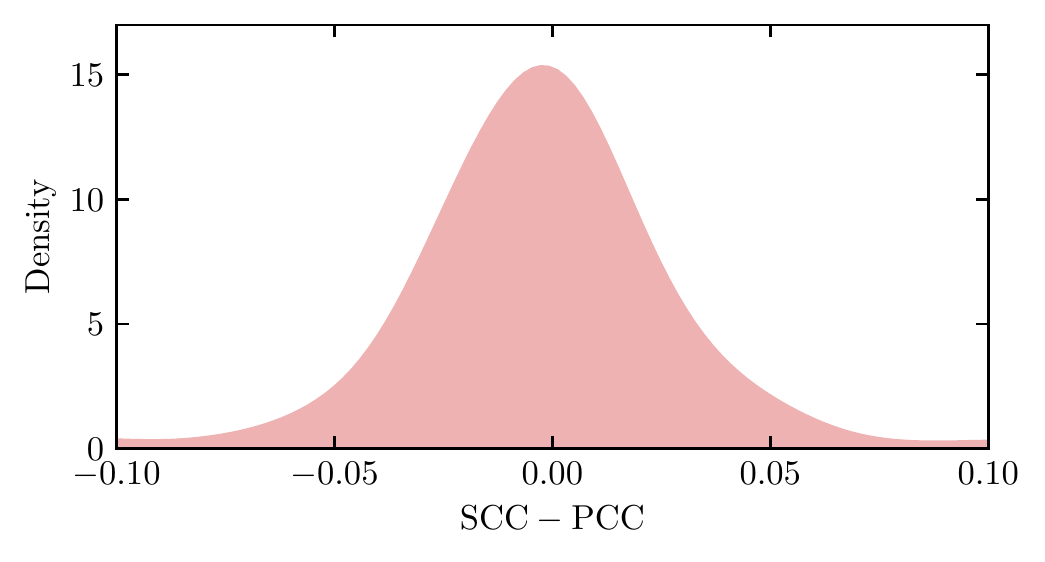}\vspace{-0.3cm}
    \caption{Difference between SCCs and PCCs for all the relations considered in the tables above.}\vspace{-0.5cm}
    \label{fig:SCC-PCC}
\end{figure}

Finally, we adopt the SCC as it does not assume a specific function form for the relationships between the different parameters. Nevertheless, we also have computed the Pearson correlation coefficients (PCCs), which assume linear relations. Figure~\ref{fig:SCC-PCC} shows the distribution of the difference between the SCCs and PCCs corresponding to the cases in all the above tables (total of 183 relations). The distribution is centered in zero and is consistent with small differences between SCCs and PCCs, which indicates that the assumption made by the multivariate regression is still valid. \vspace{-0.5cm}

\begin{table}\small\vspace{-0.3cm}
    \centering
    \begin{tabular}{r|c|c}
    \multicolumn{3}{c}{\avsph }\\\hline\hline
    
    & $\text{T}_{\text{eff}\odot}\pm 100$ K & $T_\text{eff, Doris}\pm 100$ K \\
    & $\text{L}_\odot\pm 0.3$ & $L_\text{Doris}\pm 0.3$ \\\hline
    $c_0$ & 3.98 & 4.18\\
    $c_{T_\text{eff}}$ & $3.0\times10^{-4}$ & $1.8\times10^{-4}$\\
    $c_{L}$ & $-9.7\times10^{-1}$ & $-1.0\times10^{0}$\\
    $c_\text{[Fe/H]}$ & $6.6\times10^{-1}$ & $5.5\times10^{-1}$\\
    $c_{\log\,P_\text{rot}}$ & $-1.6\times10^{0}$ & $-1.2\times10^{0}$\\\hline
    
    \multicolumn{3}{c}{}\\

    \multicolumn{3}{c}{\stdsph}\\\hline\hline
    $c_0$ & 1.27 & 0.67\\
    $c_{T_\text{eff}}$ & $-1.1\times10^{-4}$ & $-3.3\times10^{-5}$\\
    $c_{L}$ & $-1.2\times10^{-1}$ & $-6.6\times10^{-2}$\\
    $c_\text{[Fe/H]}$ & $4.0\times10^{-3}$ & $1.8\times10^{-2}$\\
    $c_{P_\text{rot}}$ & $-9.8\times10^{-3}$ & $-8.8\times10^{-3}$\\
    $c_{\log\,S\!_\text{ph}}$ & $7.6\times10^{-1}$ & $7.8\times10^{-1}$ \\\hline\hline

    \end{tabular}
    \caption{Regression coefficients for all Sun- and Doris-like stars (concerning Tables~\ref{tab:sunlike} and \ref{tab:dorislike}).}
    \label{tab:dorislike_coeff}\vspace{-0.5cm}
\end{table}\normalsize

\begin{table}\small
    \centering
    \begin{tabular}{r|c|c}
    \multicolumn{3}{c}{\avsph }\\\hline\hline
    
    & $\text{T}_{\text{eff}\odot}\pm 100$ K & $T_\text{eff, Doris}\pm 100$ K \\
    & $\log\,\text{g}_\odot\pm 0.1$ dex & $\log\,g_\text{Doris}\pm 0.1$ dex \\\hline
    $c_0$ & 6.77 & 4.01\\
    $c_{T_\text{eff}}$ & $-1.5\times10^{-4}$ & $2.3\times10^{-4}$\\
    $c_{L}$ & $-6.7\times10^{-1}$ & $-9.0\times10^{-1}$\\
    $c_\text{[Fe/H]}$ & $9/0\times10^{-1}$ & $7.0\times10^{-1}$\\
    $c_{\log\,P_\text{rot}}$ & $-2.1\times10^{0}$ & $-1.4\times10^{0}$\\\hline

    \multicolumn{3}{c}{}\\
    
    \multicolumn{3}{c}{\stdsph}\\\hline\hline
    $c_0$ & 1.43 & 0.49\\
    $c_{T_\text{eff}}$ & $-1.5\times10^{-4}$ & $-1.9\times10^{-5}$\\
    $c_{L}$ & $-1.4\times10^{-1}$ & $-9.4\times10^{-2}$\\
    $c_\text{[Fe/H]}$ & $6.8\times10^{-2}$ & $3.8\times10^{-2}$\\
    $c_{P_\text{rot}}$ & $-9.9\times10^{-3}$ & $-7.3\times10^{-3}$\\
    $c_{\log\,\langle S\!_\text{ph}\rangle}$ & $7.8\times10^{-1}$ & $8.1\times10^{-1}$ \\\hline\hline

    \end{tabular}
    \caption{Same as in Table~\ref{tab:dorislike_coeff} but only for the stars with spectroscopic constraints.}
    \label{tab:dorislike_coeff_spec}\vspace{-1.3cm}
\end{table}\normalsize

\begin{table*}\small
    \centering
    \begin{tabular}{rr|cccc|c}
    \hline\hline
    & & \multicolumn{4}{c|}{main sequence} & \\
    & & M & K & G & F & subgiant \\\hline
    
    \multicolumn{2}{r|}{N stars} & 308 & 10,850 & 16,306 & 6,117 & 2,117\\
    \multicolumn{2}{r|}{$\langle S_\text{\!ph}\rangle_{5^\text{th}}$} & 630.37 & 287.1 & 83.6 & 29.3 & 37.7\\
    \multicolumn{2}{r|}{$\langle S_\text{\!ph}\rangle_{95^\text{th}}$} & 5776.2 & 5211.4 & 4994.5 & 2118.3 & 4624.4\\\hline
    
    \multirow{7}{*}{\rotatebox[origin=c]{90}{SCC}} & Kp & 0.03 & 0.07 & 0.14 & 0.30 & 0.19 \\
    & \tobs & 0.01 & -0.10 & -0.12 & -0.20 & -0.18 \\
    & \teff & 0.12 & 0.10 & -0.07 & -0.16 & -0.23 \\
    & $L$ & -0.12 & 0.00 & -0.10 & -0.16 & 0.05 \\
    & $\text{[Fe/H]}$ & -0.04 & 0.05 & 0.00 & -0.01 & -0.02 \\
    & \prot & -0.33 & -0.36 & -0.37 & -0.17 & -0.16 \\
    & \avsph & 0.82 & 0.91 & 0.94 & 0.88 & 0.88 \\\hline

    \multicolumn{2}{r|}{$c_0$} & -1.39 & 0.16 & 0.73 & 1.34 & 1.06\\
    \multicolumn{2}{r|}{$c_\text{Kp}$} & $4.2\times10^{-3}$ & $1.3\times10^{-2}$ & $2.5\times10^{-2}$ & $5.5\times10^{-2}$ & $4.1\times10^{-2}$\\
    \multicolumn{2}{r|}{$c_{t_\text{obs}}$} & $-3.6\times10^{-3}$ & $-1.6\times10^{-2}$ & $-2.4\times10^{-2}$ & $-4.8\times10^{-2}$ & $-4.5\times10^{-2}$\\
    \multicolumn{2}{r|}{$c_{T_\text{eff}}$} & $3.8\times10^{-4}$ & $3.6\times10^{-5}$ & $-4.4\times10^{-5}$ & $-1.4\times10^{-4}$ & $-1.2\times10^{-4}$ \\
    \multicolumn{2}{r|}{$c_{L}$} & $-2.0\times10^{0}$ & $-9.4\times10^{-3}$ & $-1.9\times10^{-2}$ & $-1.6\times10^{-2}$ & $1.4\times10^{-3}$ \\
    \multicolumn{2}{r|}{$c_\text{[Fe/H]}$} & $-3.1\times10^{-2}$ & $3.7\times10^{-2}$ & $4.9\times10^{-3}$ & $-8.8\times10^{-2}$ & $-3.3\times10^{-2}$ \\
    \multicolumn{2}{r|}{$c_{P_\text{rot}}$} & $-4.3\times10^{-3}$ & $-6.2\times10^{-3}$ & $-8.4\times10^{-3}$ & $-5.3\times10^{-3}$ & $-2.2\times10^{-3}$\\
    \multicolumn{2}{r|}{$c_{\log\,\langle S_\text{ph}\rangle}$} & $9.1\times10^{-1}$ & $8.2\times10^{-1}$ & $7.7\times10^{-1}$ & $7.1\times10^{-1}$ & $7.8\times10^{-1}$ \\\hline\hline
    \end{tabular}
    \caption{Same as in Tables~\ref{tab:residuals} and \ref{tab:residuals-coefficients}, but when neglecting targets with RUWE>1.2.}
    \label{tab:ruweleq}
\end{table*}\normalsize

\begin{table*}\small
    \centering
    \begin{tabular}{rr|cccc|c}
    \hline\hline
    & & \multicolumn{4}{c|}{main sequence} & \\
    & & M & K & G & F & subgiant \\\hline
    
    \multicolumn{2}{r|}{N stars} & 31 & 2,946 & 7,767 & 4,507 & 1,629\\
    \multicolumn{2}{r|}{$\langle S_\text{\!ph}\rangle_{5^\text{th}}$} & 902.9 & 266.1 & 63.6 & 28.1 & 33.1\\
    \multicolumn{2}{r|}{$\langle S_\text{\!ph}\rangle_{95^\text{th}}$} & 8504.9 & 5346.3 & 4880.7 & 1571.7 & 3247.0\\\hline
    
    \multirow{7}{*}{\rotatebox[origin=c]{90}{SCC}} & Kp & -0.02 & 0.00 & 0.04 & 0.18 & 0.10 \\
    & \tobs & 0.02 & -0.11 & -0.14 & -0.20 & -0.15 \\
    & \teff & 0.24 & 0.06 & -0.10 & -0.15 & -0.26 \\
    & $L$ & -0.33 & 0.00 & -0.16 & -0.19 & 0.04 \\
    & $\text{[Fe/H]}$ & 0.10 & 0.02 & 0.02 & -0.06 & -0.05 \\
    & \prot & -0.43 & -0.37 & -0.36 & -0.13 & -0.11 \\
    & \avsph & 0.85 & 0.92 & 0.95 & 0.88 & 0.87 \\\hline

    \multicolumn{2}{r|}{$c_0$} & -0.86 & 0.51 & 1.23 & 1.75 & 1.48\\
    \multicolumn{2}{r|}{$c_\text{Kp}$} & $-5.2\times10^{-2}$ & $1.4\times10^{-3}$ & $1.0\times10^{-2}$ & $4.9\times10^{-2}$ & $3.1\times10^{-2}$\\
    \multicolumn{2}{r|}{$c_{t_\text{obs}}$} & $1.1\times10^{-2}$ & $-3.1\times10^{-2}$ & $-4.3\times10^{-2}$ & $-6.9\times10^{-2}$ & $-5.7\times10^{-2}$\\
    \multicolumn{2}{r|}{$c_{T_\text{eff}}$} & $4.2\times10^{-4}$ & $2.3\times10^{-5}$ & $-6.0\times10^{-5}$ & $-1.2\times10^{-4}$ & $-1.4\times10^{-4}$ \\
    \multicolumn{2}{r|}{$c_{L}$} & $-3.7\times10^{0}$ & $-4.0\times10^{-3}$ & $-2.8\times10^{-2}$ & $-1.6\times10^{-2}$ & $3.2\times10^{-4}$ \\
    \multicolumn{2}{r|}{$c_\text{[Fe/H]}$} & $1.9\times10^{-1}$ & $4.1\times10^{-2}$ & $1.2\times10^{-2}$ & $-7.9\times10^{-2}$ & $-5.7\times10^{-2}$ \\
    \multicolumn{2}{r|}{$c_{P_\text{rot}}$} & $-4.7\times10^{-3}$ & $-6.0\times10^{-3}$ & $-7.8\times10^{-3}$ & $-4.2\times10^{-3}$ & $-1.7\times10^{-3}$\\
    \multicolumn{2}{r|}{$c_{\log\,\langle S_\text{ph}\rangle}$} & $9.1\times10^{-1}$ & $8.5\times10^{-1}$ & $8.1\times10^{-1}$ & $7.2\times10^{-1}$ & $7.8\times10^{-1}$ \\\hline\hline

    \end{tabular}
    \caption{Same as in Tables~\ref{tab:residuals} and \ref{tab:residuals-coefficients}, but when considering only the targets with spectroscopic atmospheric parameters.}
    \label{tab:spec}
\end{table*}

\end{document}